\documentclass[11pt]{article}
\usepackage[letterpaper,includefoot,top=1in,bottom=1in,left=1in,right=1in]{geometry}
\usepackage{multirow}
\usepackage{float}
\usepackage[centertags,sumlimits]{amsmath}
\usepackage{amsfonts}
\usepackage{amssymb,graphics,psfrag}
\usepackage{array,epsfig,multirow,stmaryrd,graphicx}
\usepackage[all,cmtip]{xy}
\usepackage{mathrsfs}
\usepackage{color}
\usepackage{arydshln}  
\usepackage[bookmarks,breaklinks]{hyperref}
\usepackage{multirow}
\usepackage{eufrak}
\usepackage{verbatim}
\numberwithin{equation}{section}
\numberwithin{table}{section}
\usepackage{caption}
\usepackage{subcaption}
\usepackage{float}

\usepackage{bbm, dsfont}

\newcommand{\pid}{(2 \pi)^3} 

\DeclareMathAlphabet{\mathbbold}{U}{bbold}{m}{n}

\newcommand{\pa}{\partial}

\definecolor{rojo}{rgb}{0.8,0.1,0}

\renewcommand{\Im}{\operatorname{Im}}
\renewcommand{\Re}{\operatorname{Re}}

\def\H{\mathscr{H}}
\def\F{\mathscr{F}}

\def\K{\mathscr{K}}

\newcommand{\beq}{\begin{equation}}
\newcommand{\eeq}{\end{equation}}
\newcommand{\be}{\begin{equation}}
\newcommand{\ee}{\end{equation}}
\newcommand{\bea}{\begin{eqnarray}}
\newcommand{\eea}{\end{eqnarray}}
\newcommand{\ben}{\begin{eqnarray*}}
\newcommand{\een}{\end{eqnarray*}}
\newcommand{\ba}{\begin{aligned}}
\newcommand{\ea}{\end{aligned}}
\newcommand{\bt}{\begin{tabular}}
\newcommand{\et}{\end{tabular}}
\newcommand{\bc}{\begin{center}}
\newcommand{\ec}{\end{center}}

\newcommand{\nn}{\nonumber}

\newcommand{\cref}{{\bf [check ref]}}

\usepackage{amsmath}
\graphicspath{{./Pictures/}}

\entrymodifiers={+!!<0pt,\fontdimen22\textfont2>}

\begin{document}

\baselineskip=16pt
\setlength{\parskip}{6pt}

\begin{titlepage}

\begin{center}

{\Large \bf Mirror quintic vacua: hierarchies and inflation}

\vskip 1cm

\begin{center}

{ \bf Nana Cabo Bizet}$^{a-b,}$\footnote{\texttt{nana@fisica.ugto.mx}},
{\bf Oscar Loaiza-Brito}$^{b,}$\footnote{\texttt{oloaiza@fisica.ugto.mx}},
{ \bf Ivonne Zavala}$^{c,}$\footnote{\texttt{e.i.zavalacarrasco@swansea.ac.uk}}
 \end{center}
\vskip 0.0cm
 \emph{$^{a}$ Mandelstam Institute for Theoretical Physics, School of Physics, \\
 and NITheP, University of the Witwatersrand, WITS 2050, Johannesburg, South Africa} 
\\[0.1cm]
 \emph{$^{b}$ Departamento de F\'{\i}sica,  Universidad de Guanajuato,\\
 Loma del Bosque 103, CP 37150,  Le\'on, Guanajuato, M\'exico} 
  \\[.1cm]
\emph{$^{c}$ Department of Physics, Swansea University, Singleton Park, Swansea, SA2 8PP, UK}

\end{center}
\vskip 0.0cm

\begin{center} {\bf ABSTRACT }
\end{center}

We study the moduli space of type IIB string theory flux compactifications  on the mirror of the CY quintic 3-fold in $\mathbb{P}^4$. We focus on the dynamics of the four dimensional moduli space, defined by the axio-dilaton $\tau$ and the complex structure modulus $z$. The $z$-plane has critical points, the conifold, the orbifold and the large complex structure with non trivial monodromies. We find the solutions to the Picard-Fuchs equations obeyed by the periods of the CY in the full $z$-plane as a series expansion in $z$ around the critical points to arbitrary order. This allows us to discard fake vacua, which appear as a result of keeping only the leading order term in the series expansions. Due to monodromies vacua are located at a given sheet in the $z$-plane.
A dS vacuum appears for a set of fluxes. We revisit vacua with hierarchies among the 4D and 6D  physical scales close to the conifold point  and compare them with those found at leading order in \cite{Giddings:2001yu,Ahlqvist:2010ki}. 
We explore slow-roll inflationary directions of the scalar potential by looking at regions where the multi-field slow-roll parameters $\epsilon$ and $\eta$ are smaller than one. The value of $\epsilon$ depends strongly on the approximation of the periods and to achieve a stable value, several orders in the expansion are needed. We do not find realisations of single field axion monodromy inflation. Instead, we find that  inflationary regions appear along linear combinations of the four real field directions and for certain configurations of fluxes.\\

\hfill \today
\end{titlepage}

\tableofcontents

\section{Introduction}

Type IIB string theory flux compactifications  on Calabi-Yau (CY) orientifolds have become a fruitful scenario to construct effective four dimensional models with desirable phenomenological features. These compactifications  attracted a lot of attention after the seminal work of Giddings, Kachru and Polchinski (GKP) \cite{Giddings:2001yu}, where  a mechanism for stabilisation of the axio-dilaton and the complex structure moduli was found based on the
flux superpotential \cite{Gukov:1999ya,Curio:2000sc}. Moreover, these compactfications provide a rich arena to study phenomenology. Fluxes' backreaction causes the internal CY manifold to be highly warped, as studied in  \cite{KS}  where uncompact  flux supergravity solutions near the conifold were found. This opened up the possibility that large hierarchies among the  space-time and compactification physical scales can  be realised in string theory constructions as discussed in \cite{Giddings:2001yu}. 
Using this approach to moduli stabilisation,  extensive studies appeared on the possibility to construct effective 4D models with (meta)stable de Sitter vacua, starting with the work of \cite{KKLT} and/or with  regions of moduli space suitable for slow-roll inflation.

Of particular interest are stringy models of inflation with potentially detectable primordial gravitational waves\footnote{Although whether primordial gravitational waves can be realised within perturbative string theory remains an open question \cite{KPZ}.} \cite{Silverstein:2008sg,AM2,FI,AZ}. These models predict, via the Lyth bound \cite{Lyth}, a super-Planckian field excursion and  are thus  particularly sensitive to ultraviolet corrections  through  higher dimensional operators and by quantum corrections to the inflaton mass.  Therefore  a mechanism to prevent such corrections while preserving the inflationary conditions is required.  An attractive possibility to address this issue  is to invoke  a symmetry that forbids large quantum corrections.  In particular,  the shift symmetry governing axions    can provide such a symmetry. The continuous shift symmetry can be broken by non-perturbative effects to a discrete one, as in natural inflation \cite{NI}, or spontaneously due to couplings to non-trivial background flux for example, giving rise to a realisation of monomial or chaotic inflation \cite{Linde} with axions \cite{KaSo}.  String theoretic embeddings of the later possibility are  known as axion monodromy inflation \cite{Silverstein:2008sg,AM2}.   

Of particular interest for our work is F-term axion monodromy, which arises through the F-term in the superpotential  \cite{Marchesano:2014mla,Blumenhagen:2014gta,Blumenhagen:2014nba,Hebecker:2014kva}. 
One interesting possibility to realise F-term axion monodromy inflation in string theory\footnote{For further recent studies on axion monodromy see \cite{Escobar:2015fda,Andriot:2015aza,Escobar:2015ckf,Hebecker:2015tzo}  and in non-geometric compactifications  \cite{Blumenhagen:2015kja,Blumenhagen:2015jva,Blumenhagen:2015xpa}.} is to use the axion directions of the complex structure moduli \cite{Blumenhagen:2014nba,Hebecker:2014kva,Hebecker:2015rya,Tatsuo} .  
To study the complex structure (CS) moduli space of the internal six dimensional manifold in CY compactifications, one has to study  the periods of the CY.  These are defined as the integrals of the holomorphic $(3,0)$-form over the 3-cycles of the manifold in an integral symplectic basis of $H_3(CY,\mathbb{Z})$.  
According to the theorem of Landman \cite{landman},  at certain critical points in the complex structure moduli space, the periods can have a logarithmic or finite order branch cut behaviour. This leads to a monodromy  matrix $\mu_i \in Sp(b_3,\mathbb{Z})$ that acts on the periods $\Pi(z)$  when the i-th critical point is encircled. The matrix $\mu_i$ has the property $(\mu_i^k-{\bf{1}})^{p+1}=0, \ \ {\rm with} \ p\le 3 \ $ 
where $p$ is the smallest integer so that the r.h.s. is zero.  For $p=0$, $k>1$ there is
an $\mathbb{Z}_k$ orbifold singularity. The cases  $k=1$ are the unipotent cases. The conifold has $p=1$ and the maximal unipotent case is $p=3$. 
In the case of the mirror quintic one has a conifold point, a  large complex structure (LCS) point, or maximal unipotent point and a $\mathbb{Z}_5^3$ orbifold point. 

Hence the monodromies  around each of the different critical points  are of different nature. 
For example, the monodromy around the orbifold has a finite order. Instead, monodromies around the LCS and conifold points have infinite order. This has triggered interest in these critical points as potential set ups  to realise large field inflation in string theory supergravity models, using the mechanism of axion monodromy \cite{Silverstein:2008sg,AM2}. 
Indeed, the K\"ahler potential is invariant under  a shift symmetry, or more generally a monodromy along the argument of the  complex structure, $\arg(z)\rightarrow \arg(z)+2\pi n$,  which is broken spontaneously by the fluxes. 

A common simplification  used in  the literature  to realise CS axion monodromy in these type of compactifications is to consider only the leading order term in the series expansion of the periods in terms of the CS moduli \cite{Blumenhagen:2014nba,Hebecker:2015tzo,Hebecker:2015zss, Garcia-Etxebarria:2014wla} 
when computing the scalar potential
$$V = \frac{e^K }{2\kappa_{10}^2 g_s}\left(|DW|^2 -3|W|^2\right),$$ 
where the periods enter into the superpotential $W$ and the K\"ahler potential $K$ (see below). 
 However, one may worry that by cutting  the series at the leading order might lead to apparent vacua,  which disappear  when higher order terms are included or  miss interesting regions for cosmological applications. In \cite{Tatsuo} for example, the authors  kept up to second order in the periods' expansion and they found interesting new potentials suitable to study natural inflation. 

In the present work we consider  this problem in detail by studying  type IIB orientifold flux compactifications keeping all orders in the series  expansion of the periods in the complex structure modulus, necessary to achieve convergence of the  solutions we study. To achieve this, we need to keep up to order 600 in the series' expansion. In this sense our computations are exact. 
We focus on the CY  mirror of the quintic on $\mathbb{P}^4$ \cite{Candelas:1990rm} and 
 compute the periods of this manifold  in four different patches, around the three  singular points, and around a regular point, keeping all necessary orders in the series until  convergence is  achieved.   
   We also obtain the transition functions that allow us to move from one patch to the other,  covering   the whole CS moduli space.

The  CS moduli space of the mirror quintic has  been previously studied in \cite{Candelas:1990rm,Huang:2006hq}.
We study the periods around the conifold as in \cite{Huang:2006hq}, but do this up  to a higher order in the series expansion in $z$, until  convergence is achieved in the search for vacua and inflationary regions.  We also compute the periods around the orbifold,  the LCS and around regular points in the $z$-plane. The solutions around regular points serve to explore regions close to the boundaries of convergence around the critical points. To date, the periods around the conifold in the integer symplectic basis are only known numerically. This is because the transition matrices  (conifold-orbifold, conifold-LCS) can  only be determined numerically. Considering all these patches and the transition matrices  we study the whole complex structure space in the mirror of the quintic on $\mathbb{P}^4$. This method has been used in \cite{Bizet:2014uua} to study the complex structure space of the 4-fold mirror of the sextic on $\mathbb{P}^5$.

Including the higher order terms in the series until convergence is achieved,  we accomplish various goals. First, we find global (no-scale) vacua appearing beyond the leading order in $z$. That is, by  obtaining the integer symplectic basis in the whole CS moduli space, we are able to explore the vacua landscape in regions far from the critical points.  We verify the existence of hierarchies in the physical scales as found in \cite{Giddings:2001yu}, where they took only the leading order contributions to the periods, $\Pi(z)$. 
We  find  a  correction of order one to this result.  We also find that in general, the exact vacua  (i.e.~the solutions found keeping all orders in the series until convergence is reached), differ from the near conifold approximation solutions studied in  \cite{Giddings:2001yu} and \cite{Ahlqvist:2010ki}.  We verify that hierarchies are a generic feature of flux compactifications.  
Finally, we look for inflationary regions, where the multi-field slow-roll parameters are smaller than unity, by varying the fluxes and moving through the whole moduli space. We find small slow-roll parameters over large moduli regions. These  inflationary regions seem to happen generically in a  multi-field fashion, such that during inflation there are field's displacements along all of the moduli directions. 

As mentioned already, we are interested in the dynamics and properties of the complex structure $z$, and axio-dilaton $\tau$ moduli space. Therefore, we focus on no-scale models of the mirror quintic.  
This will allow us to explore the CS  and the dilaton moduli space in more detail and to identify whether or not the axions associated to these fields can be used as inflaton candidates. In order to 
stabilize the  K\"ahler moduli, it would be necessary to include non-perturbative contributions to the superpotential\footnote{For example, in \cite{DDFGK} parametrically controlled moduli stabilisation of the $h^{1,1}=51$ K\"ahler moduli was demonstrated extending the analysis to F-theory \cite{Bizet:2014uua}.} or to add (non)-geometric fluxes. 
It is interesting  to mention that a  large number of K\"ahler moduli could increase the chances to find axionic inflationary regions as recently studied in \cite{BaMa,Long:2016jvd}.

The paper is structured as follows. In Section \ref{fluxC} we fix our notation and conventions by reviewing orientifold flux compactifications in type IIB string theory. In \ref{sugra} we write down the relevant  ${\cal N}=1$ supergravity action in four dimensions and we describe the integral symplectic basis for the periods. In Section~\ref{mirrorQu} we discuss the properties of the mirror quintic CY. Finally in Section~\ref{PFeq} we  describe the procedure to solve the PF equations for the periods in all different patches (orbifold-, LCS-, conifold- and a regular point convergence regions)
giving explicitly approximated expressions. 
We finish Section \ref{fluxC} by describing  the monodromies as shift symmetries of the K\"ahler potential 
and their breaking by the flux generated superpotential  in \ref{symmetries}.
In Section \ref{mirrorQ} we focus on the scalar potential 
and describe the  vacua that we find for different flux configurations. In Section~\ref{hierarchies} we study the vacua near the conifold
in \cite{Giddings:2001yu,Ahlqvist:2010ki}  and compare them with the exact vacua that we find\footnote{Note again that we refer to the vacua we find taking into account all necessary terms in the series for the periods needed for the solutions to converge. }.
In Section~\ref{sVacua} we describe in detail the  vacua we find inside the conifold convergence region.  Finally in \ref{inflation} we explore inflationary regions,  where the multi-field slow-roll parameters are small for large regions of the moduli space. 
We find inflationary regions extending from the conifold- to the orbifold point where
no particular direction in the moduli space is favoured. That is, inflation seems to occur generically in a multi-field fashion. In particular, the effective inflationary direction does not seem to occur particularly along any shift symmetric directions ($\arg(z)$ and $\Re(\tau)$). Therefore, we do not see a plausible realisation  of axion monodromy inflation in this set up. We show further  that the $\epsilon$ parameter is very sensitive to the approximation considered for the period's series expansion. 
We conclude the main text in Section \ref{conclusions} with a discussion of our findings. 
In Appendix \ref{transition1}   we discuss  the integral symplectic basis for the periods  in the different patches and present  the relevant  transition functions.  In Appendix \ref{Ahierar} a correction of one order of magnitude is given to the hierarchy formula of \cite{Giddings:2001yu}.  
Finally in Appendix \ref{appC}  we give an analytical  description of the scalar potential along a SUSY preserving direction for the axio-dilaton ($D_\tau W=0$), moving with monodromies  along the conifold point.

\section{Type IIB flux compactification on the  mirror quintic}
\label{fluxC}
In this section we review the basic ingredients of  four dimensional supergravity which arises from the low energy limit of type IIB string theory compactified on Calabi-Yau (CY) orientifolds with non-trivial RR and NS-NS 3-form fluxes. We describe the integral symplectic basis for the CY periods, 
which is required for flux quantization. Along the way we fix our notation and conventions. 
\subsection{${\cal N} =1 $ SUGRA}
\label{sugra}

We start with the ten dimensional effective supergravity action in the Einstein frame\footnote{We use the conventions for transforming to the Einstein frame $G_{MN}^E=e^{(\phi-\phi_0)/2}G_{MN}^s$, where $G_{MN}$ is the 10D metric, $\langle e^{\phi}\rangle = e^{\phi_0}=g_s$ with $\phi$ the dilaton and $g_s$ is the string coupling. With these conventions the volumes are conformally invariant.} including fluxes and focus on the effective four dimensional action after dimensionally reducing it  (see \cite{Giddings:2001yu} for details). That is, in four dimensions we are interested in the action  
\be\label{4daction}
S_4 = \int{ d^4 x \sqrt{g } \left[ \frac{M_{Pl}^2}{2} R  - M_{Pl}^2 \, K_{a\bar b}\, \partial_\mu \Phi^a \partial^\mu \bar \Phi^{\bar b} + V(\Phi^l)\right] }\,,
\ee
where  $M_{Pl}^2 = 1/\kappa_4^2= V_6/(\kappa_{10}^2 g_s^2)$ is the  Planck scale,  $V_6$ is the dimensionfull 6D volume$, \kappa_{10}^2= (2\pi)^7(\alpha')^4/2 \equiv \ell_s^8/4\pi$,  $\ell_s=\sqrt{2\pi} \alpha'$ the string scale and $g_s=\langle e^{\phi}\rangle$ the string coupling. The indices $a, b$ run over the moduli fields present, which are the axio-dilaton $\tau = C_0 +i \,e^\phi$, the complex structure $z_i$, $i=1,\dots , h^{2,1}$ and the K\"ahler moduli, $T_m$, $m=1,\dots, h^{1,1}$.  
The K\"ahler potential for the moduli is given by
\be\label{kahler}
K= -\ln \left[-i \left(\tau-\bar \tau \right)\right] -\ln \left[i \int_{CY} \Omega\wedge \bar \Omega\right] - 2\ln \left[{\cal V} \right]\,,
\ee
where $\Omega$ is the holomorphic $(3,0)$ form of the CY  and
 $\cal V$ is the dimensionless volume defined in terms of the dimensionless K\"ahler moduli $T_m$.

 The complex structure moduli can be parameterised  by the integrals of  $\Omega$ over a canonical homology basis of the CY. These are known as the periods, $\Pi$ of the CY. In this work according to \cite{Candelas:1990qd} we use the canonical integral symplectic basis $(\alpha_I,\beta^I)$ on $H^3(CY,\mathbb{Z})$  and its dual homology basis $(A^I,B_I)$ of $H_3(CY,\mathbb{Z})$  satisfying
\begin{equation}
\int_{CY} \alpha_I\wedge \beta^J=\delta_I^J = - \int_{CY}  \beta^J \wedge \alpha_I\,, \qquad  
\int_{CY} \alpha_I\wedge \alpha_J= \int_{CY}  \beta^I \wedge \beta^J=0 \,,
\end{equation}
\begin{equation}
\int_{A^J} \alpha_I=- \int_{B_I} \beta^J=\delta_I^J  \,.
\end{equation}

The indices $I$ and $J$ run from $0$ to $h^{2,1}$.  With respect to this basis the CY periods  are defined as
\begin{equation} \label{periods1}
\Pi=\binom{\mathcal{X}^I}{\mathcal{F}_I}=\left(
\begin{array}{c}
\int_{A^I} \Omega \\ \int_{B_I} \Omega
\end{array}\right)  \,,
\end{equation}
and consequently, the holomorphic 3-form can be expanded as 
\begin{eqnarray}
\Omega=\mathcal{X}^I\alpha_I-\mathcal{F}_I\beta^I\,.\label{omeg}
\end{eqnarray}
Similarly the K\"ahler potential for the complex structure moduli is given by  
\be
K_{CS} = -\ln\left(-i\, \bar \Pi^T\,\Sigma \,\Pi \right),
\label{K}
\ee
where $\Sigma$ denotes the symplectic matrix, defined as
\begin{equation*}
\Sigma=\left(
\begin{array}{cc}
0&\mathbbm{1}_{k\times k}\\
-\mathbbm{1}_{k\times k}&0\
\end{array}\right),
\end{equation*}
with $k=1+h^{2,1}$.

In the following we shortly review the integral symplectic basis (\ref{periods1}) which is required for flux quantization.
This basis is the one employed through the paper in all the different patches in the CS moduli space.
Special geometry implies the existence of a holomorphic prepotential ${\cal F}$, which is homogeneous of degree two 
in the ${\cal X}^I$. The ${\cal F}_I$ are given as derivatives  ${\cal F}_I=\frac{\partial {\cal F}}{\partial {\cal X}_I}$. 
The prepotential determines the periods, the couplings, as well as the K\"ahler potential, see e.g. \cite{Candelas:1990pi}.  
Mirror symmetry implies that at the large radius point of a CY 3-fold $M_3$, corresponding to the
large complex structure (LCS) point on the mirror $W_3$ the prepotential reads as follows \cite{Candelas:1990pi,Hosono:1994ax}
  \begin{eqnarray}
{ \cal F}
&=&-{C^0_{ijk} {\cal X}^i {\cal  X}^j {\cal X}^k \over 3! {\cal X}^0}+ n_{ij} {{\cal X}^i {\cal X}^j \over 2}+ c_i {\cal X}^i {\cal X}^0-i{\chi 
\zeta(3)\over 2 \pid}({\cal X}^0)^2+ ({\cal X}^0)^2 f(q)\nonumber\\
&=& ({\cal X}^0)^2{\widetilde{\cal F}}= ({\cal X}^0)^2\left[-{C^0_{ijk} t^i t^j t^k\over 3!}
+n_{ij} {t^i t^j \over 2}+
c_i t^i-i{\chi \zeta(3)\over 2 \pid} +f(q)\right], \label{geomprep}
\end{eqnarray}
where $i,j,k=1,\ldots,h^{2,1}$, $q_i=\exp (2 \pi i t_i)$, $f(q)$ represents the instanton contributions, $C^0_{ijk},c_{ij}$, $n_i$ and $\chi$ are topological data of the manifold \cite{Bizet:2014uua}.  The integral basis for the periods at the LCS point
is then given by
\beq
\Pi_{LCS}=\left(\begin{array}{c} 
\mathcal{X}^0  \\ 
\mathcal{X}^i \\
{\mathcal{F}_0 }\\
{\mathcal{F}_i}
\end{array}\right)=\mathcal{X}^0\left(\begin{array}{c} 
1   \\
t^i \\
2 \widetilde{{\cal F}}- t^i  \pa_i \widetilde{{\cal F }}\\
{\pa {\widetilde{\cal F}}\over \pa t^i}\end{array}\right)={\cal X}^0\left(\begin{array}{c}
1\\
t^i\\
{C^0_{ijk}\over 3!} t^i t^j t^k+c_i t^i-i{\chi \zeta(3)\over \pid}
+f(q)\\
-{C^0_{ijk}\over 2} t^i t^j+{n_{ij}} t^j+c_i+\pa_i f(q)
\end{array}\right)\ .
\label{Pi1lcs} 
\eeq
The mirror map reads $t^i=\frac{{\cal X}^i}{{\cal X}^0}=\frac{1}{2 \pi i} \left(\log(z_i)+ \Sigma^i({\underline z})\right)$, $i=1,\ldots, h^{2,1}$,
where $z_i$ are the complex structure moduli and $\Sigma^i(\underline z)$ are power series in $z_i$.  In Section \ref{mirrorQu} we describe such LCS point for the CY 3-fold mirror of the quintic on $\mathbb{P}^4$ which is the compactification employed. For this case $h^{2,1}=1$, $C^0_{111}=5$, $c_1=\frac{50}{24}$, $n_{11}=-\frac{11}{2}$ and $\chi(W_3)=200$. We are interested however on other critical points, in particular the conifold, close to which the periods have to be determined, and one needs to obtain their expressions in terms of the integer symplectic basis (\ref{Pi1lcs}).

Turning on fluxes in the different 3-cycles of an orientifold CY, generates a four dimensional scalar potential  for the axio-dilaton and the complex structure moduli, given by \cite{Giddings:2001yu}, 
\be\label{scalarV}
V= \frac{1}{2\kappa_{10}^2}\int_{CY}d^6y \sqrt{\tilde g}\, \,\frac{G_{(3)} \cdot \bar G_{(3)}}{12 \Im\, \tau} -\frac{i}{4\kappa_{10}^2 \Im \tau}\int_{CY} G_3 \wedge \bar G_3\,,
\ee
where the inner product of the three-form fluxes is performed using $\tilde{g}_{mn}$ given by Eq.~(\ref{metric}) below.
The first contribution comes from the fluxes and the second from the branes and orientifold charges. The 3-form flux $G_{(3)}$ is defined in terms of the RR, NS-NS flux and dilaton as:
\be
G_{(3)} = F_{(3)} -\tau H_{(3)}\,,
\ee
with
\bea
F_{(3)}=F_{(3)}^I\alpha_I-F_{(3)I}\beta^I\,, \qquad  H_{(3)}=H_{(3)}^I\alpha_I-H_{(3)I}\beta^I \,,
\eea
so that 
\be
G_{(3)}=F_{(3)} - \tau H_{(3)} =  G^I\alpha_I - G_I\beta^I\,, \\
\ee
with $G^I=F_{(3)}^I-\tau H_{(3)}^I$ and $G_I=F_{(3)I}-\tau H_{(3)I}$.

The $F_{(3)}, H_{(3)}$  fluxes on the 3-cycles of the orientifold CY  are quantised as
\bea\label{MN}
&&\frac{1}{(2\pi)^2 \alpha'} \int_{A^I} F_{(3)} = M^I \,,  \qquad  \frac{1}{(2\pi)^2 \alpha'} \int_{A^I} H_{(3)} = N^I \,, \nonumber\\
&&\frac{1}{(2\pi)^2 \alpha'} \int_{B_I} F_{(3)}= M_I \,,  \qquad  \frac{1}{(2\pi)^2 \alpha'} \int_{B_I} H_{(3)} = N_I \,.
\eea
Note that due to the Dirac quantization condition the fluxes need to be defined with respect to an integral basis of $H^3(CY,\mathbb{Z})$ given in (\ref{Pi1lcs}).
The scalar potential \eqref{scalarV} can be written in an ${\cal N} =1$ supergravity form as
\be
V =\frac{1}{2\kappa_{10}^2 g_s}\,e^K\left[K^{a\bar b}D_a W \overline{D}_{\bar b} \overline{W} -|W|^2 \right],\label{GVW}
\ee
The scalar potential (\ref{GVW}) depends on the superpotential $W$ and the K\"ahler potential.
In (\ref{GVW}) the indices $a,b$ denotes the moduli fields,  $K^{a\bar b}$  is the inverse metric in field space and $D_a W=\partial_a W+\partial_a K W$ is the supersymmetric covariant derivative of $W$. The superpotential generated by the fluxes is given by the Gukov-Vafa-Witten (GVW) superpotential \cite{Gukov:1999ya}:  
\begin{eqnarray}
\label{fluxW}
W &=& \int_{CY} G_{(3)}\wedge \Omega=   \int_{CY}( F_{(3)}- \tau  H_{(3)})  \wedge \Omega,\\
&=& (F_{(3)}^I-\tau H_{(3)}^I)\mathcal{F}_I - (F_{(3)I}-\tau H_{(3)I})\mathcal{X}^I =  
G \,\Sigma\, \Pi  \,,\nn
\end{eqnarray}
where in an abuse of notation, we have omitted the use of explicit indices in the last expression and we have defined $\int_{A^I,B_J} G_{(3)} = G = (G^I, G_J)$  where the fluxes are defined through \eqref{MN} (that is e.g.~$\int_{A^I} F_{(3)}= M^I (2\pi)^2\alpha' \equiv F_{(3)}^I$). 

We consider non-supersymmetric no-scale models \cite{Giddings:2001yu} where the the K\"ahler moduli  cancel the negative contribution to (\ref{GVW}) by setting $K^{m\bar n}D_{m} W \overline{D}_{\bar n} \overline{W} -3|W|^2=0$. Since in this case the GVW superpotential depends only on the dilaton and the complex structure moduli the scalar potential is positive definite  with the form 
\be
V =\frac{1}{2\kappa_{10}^2 g_s}\,e^K\left[K^{i\bar j}D_i W \overline{D}_{\bar j} \overline{W} \right]\,.
\ee
The indices run only over the axio-dilaton and the complex structure modulus. This is the potential that we study in the rest of the paper. As already mentioned, for our analysis we shall not consider the stabilisation of the K\"ahler moduli.
In the case of the mirror quintic $h^{2,1}=1$ and we use the following notation for the  components  of the periods and the fluxes:
\begin{eqnarray}
\Pi=\binom{\mathcal{X}^I}{\mathcal{F}_I}  
&=&\left(
\begin{array}{c}
\Pi_1\\\Pi_2\\\Pi_3\\\Pi_4\end{array}\right),  \, \\
(F_{(3)}^I,F_{(3)I})&=&(F_1,F_2,F_3,F_4),\, \nn \\
(H_{(3)}^I,H_{(3)I})&=&(H_1,H_2,H_3,H_4), \nn\\
(G_{(3)}^I,G_{(3)I})&=&(G_1,G_2,G_3,G_4). 
\end{eqnarray}\label{flujos}

\subsection{The mirror of the quintic in $\mathbb{P}^4$}
\label{mirrorQu}

We consider  the explicit orientifold compactification of type IIB string theory where the internal manifold is the mirror of the quintic hyper surface on $\mathbb{P}^4$ \cite{Candelas:1990rm}. We start by describing the properties of the CY manifold. In particular, we  describe the critical points  in the complex structure moduli space of the manifold, namely the orbifold,  large complex structure and conifold singularities and the structure of the monodromies as one encircles these critical points. We shall  describe how to compute the periods for this manifold in the vicinity of those singular points, covering thus the full complex structure moduli space. We use these results in the next section to explore new vacua and their properties, as well as potential regions to realise slow-roll inflation.
Let us first  review  the construction of the mirror of the quintic CY in $\mathbb{P}^4$ following \cite{Candelas:1990rm}. The quintic CY is the 3-fold constructed as the most general quintic hyper surface $\tilde P=0$  in ${\mathbb P}^4$. This variety has 101  complex structure moduli corresponding to the independent coefficients entering $\tilde P$. It further has a single K\"ahler modulus and hence Euler number given by $\chi =(2 h^{1,1}-h^{2,1})=-200$. 

The mirror of the quintic CY threefold is obtained by modding out a $\mathbb{Z}_5^3$ symmetry from a one parameter family of polynomials on ${\mathbb P}^4$.
This family is given by
\be
W_{\psi}=\left \{(x_1,x_2,x_3,x_4,x_5)\in \mathbb{P}^4, \, P= \sum_{k=1}^5 x_k^5 - 5 \psi \prod x_k =0\right \} \,. \label{mir}
\ee
$W_{\psi}$ has a $\mathbb{Z}_5^3$ symmetry generated by phase rotations $x_l\rightarrow e^{\frac{2\pi i g^{(k)}_l}{5}} x_l,\, l=1,...,5$ with $g^{(1)}=(0,1,0,0,4)$, $g^{(2)}=(0,0,1,0,4)$ and  $g^{(3)}=(0,0,0,1,4)$. The symmetry \cite{Candelas:1990rm}  is modded out to obtain the mirror quintic manifold: $W_{\psi}/\mathbb{Z}_5^3$.  
In the mirror manifold, the parameter of the invariant deformation $\psi$ constitutes the single complex structure modulus. As dictated  by mirror symmetry, the mirror quintic has $h^{1,1}=101$ K\"ahler moduli, a single complex structure modulus, $h^{2,1}=1$, which we  denote by $z=\psi^5$ and Euler number $\chi =+200$.  It also has Betti number $b_3=\sum_{i=0}^3h^{3-i,i}=4$, that is, four 3-cycles where the three-form fluxes can be turned on. Therefore it has four periods \eqref{periods1}, which are all functions of the single complex structure modulus $z$. 
There are three critical points in the complex structure moduli space, the orbifold, the conifold and the large complex structure. Both the conifold and the large complex structure points arise when 
\be
P=dP=0\,.\label{dP}
\ee
The conifold arises at the locus $ \forall_i\, 5 x_i^5-5\psi x_1 x_2 x_3 x_4 x_5=0$, which is  satisfied for 
$\psi^5=1$ and $|x_i|=1$. At this point the CY has a nodal singularity. The modulus $\psi$ can be parametrized by the coordinate
$z_C$ given as $z_C=1-\psi^{-5}$. At the point $\psi\to \infty$, the manifold degenerates 
to $x_1x_2x_3x_4x_5=0$. This is the  large complex structure point (LCS) also called the maximal unipotent
monodromy (MUM) point located at $z_C=1$. 
Finally the point $\psi =0$ corresponds to an specially symmetric point, the orbifold, 
located at $z_C= \infty$. Transport of the periods around the critical points $\psi_0=0,1,\infty$, lead 
to specific  monodromy transformations:
\be
\Pi \to \mu \Pi\,, 
\ee 
where $\mu$ is the monodromy transformation matrix. We denote the monodromies around the conifold, the large complex structure and the orbifold points  by $\mu_C, \mu_M$ and $\mu_O$ respectively. The monodromy around the LCS ($\psi_0 = \infty$) fulfills the condition $(\mu_M -1)^4=0$, which means that this is a point of maximal unipotent monodromy. 
The monodromy $\mu_M$ is of infinite order, which implies that at every turn around $\psi_0=\infty$ the periods acquire   different values. 
Around the orbifold  point  ($\psi_0 = 0$) the monodromy satisfies $\mu_{O}^5=1$, and it is therefore  of order 5. Finally, around the conifold point $\psi_0=1$, we have $(\mu_C -1)^2=0$ and this point is 
unipotent. Also $\mu_C$ is of infinite order. 
 The periods (\ref{periods1}) obey Picard-Fuchs (PF) equations, whose  solutions give the dependence on the complex structure modulus that will be explored in Section \ref{PFeq}. From the explicit expressions of the periods
the matrices $\mu_O,\mu_C,\mu_M$ can be obtained in any given basis, we will write them in terms of the integral symplectic basis in (\ref{Mu1}) and (\ref{MonMo}).

The mirror quintic $W_{\psi}/\mathbb{Z}_5^3$ possesses a symmetry which identifies $x_1 \leftrightarrow x_2$ leading to the
possibility of having   $O7$ planes  and $O3$-planes  \cite{Brunner:2003zm} .
A fixed hyperplane under this symmetry is given by
\begin{eqnarray}
(x_1,x_1,x_3,x_4,x_5)\in \mathbb{P}^4 \text{ with  }  2 x_1^5+ x_3^5+x_4^5+x_5^5-5\psi x_1^2x_3 x_4 x_5=0, \label{O7}
\end{eqnarray}
which has two complex internal dimensions and represents an $O7$ plane. Additionally the locus
$(1,-1,0,0,0)$, which is also fixed under $x_1 \leftrightarrow x_2$   constitutes an $O3$ plane.
That is, the $O7$ plane divisor is given by  $x_1-x_2=0$, and an $O3$ plane
appears at the point $x_1=-x_2=1,\, x_3=x_4=x_5=0$. 

Therefore the mirror quintic orientifold compactification contains O3 and O7-planes. In order to cancel tadpoles in this set-up we should also include spacetime filling D7-branes, as well as D3-branes.  We make here a short review of \cite{Blumenhagen:2008zz} applied to our setup. The 3-form fluxes  contribute to the D3-brane charge tadpole and therefore we have  %
\begin{eqnarray}
N_{flux}&=&\frac{1}{2\kappa_{10}^2 T_3}\int_{CY} F_3\wedge H_3=-\frac{g_s}{(2\pi)^4(\alpha')^2}F\cdot \Sigma \cdot H, \nn\\
&=&\frac{N_{O3}}{2}-2N_{D3}+\frac{\chi_{(D_{O7})}}{6}+\sum_a (Q^a_{D_7}+Q'^a_{D_7}),\label{tadpole1} \\
&=&\frac{N_{O3}}{2}-2N_{D3}+\frac{\chi_{(D_{O7})}}{6}+\sum_a N_a \frac{\chi(D_a)}{12}\,,\nn 
\end{eqnarray}
where $T_3 = (g_s(\alpha')^2(2\pi)^3)^{-1}$ is the D3-brane tension, the index $a$ numerates the D7 branes, $D_a$ is the divisor wrapped by the D7 brane,  $\chi(D_a)$ the corresponding Euler number,  $D_{O7}$ is the divisor that corresponds to the
O7 plane, $2N_{D3}$ is the number of D3 branes and their images.  $\chi(D_{O7})$ is computed via  \cite{Blumenhagen:2008zz}
\begin{eqnarray}
\chi(D_{O7})&=&\int_{CY} c_2(D_{O7})\wedge [D_{O7}],\label{chi}
\end{eqnarray}
with $c_2$  the second Chern class of the divisor $D_{O7}$.  
The calculation of $\chi(D_{O7})$ is  involved because of the presence of 101
K\"ahler moduli. The D7 tadpole cancellation reads  \cite{Blumenhagen:2008zz}
\begin{equation}
\sum_a N_a ([D_a]+[D'_a])=8 [D_{O7}],\label{tadpole2}
\end{equation}
where $D'_a$ is its image under the orientifold projection of $D_a$.

When the D7 branes are on top of the O7 plane we have $[D_a]=[D'_a]=[D_{O7}]$, this  simplifies the equation (\ref{tadpole2}) to $\sum_a N_a=4$.
In this case there is a zero contribution from the branes to the superpotential and the K\"ahler potential 
$W_{D7}=0$ (see (4.57) in \cite{Denef:2008wq}) and $K_{D7}=0$ (see (4.26) in \cite{Denef:2008wq}) avoiding
a mixture with CS moduli.  The D3 brane tadpole cancellation condition (\ref{tadpole1})  reduces to 
\begin{eqnarray}
N_{flux}=\frac{N_{O3}}{2}+\frac{\chi(D_{O7})}{2}-2N_{D3}.\label{tadpole3}
\end{eqnarray}

In what follows we assume for our study that by incorporating the necessary number of $D3$-branes and by 
computing the Euler number of the 
divisors in the mirror quintic, it is possible to cancel the tadpole (\ref{tadpole1}) or (\ref{tadpole3}) for any flux configuration.

\subsection{Picard-Fuchs equations}
\label{PFeq}

In this section we write the Picard-Fuchs (PF) equations, which are  fourth order differential equations, satisfied by the periods, in four different coordinates systems. Three of those correspond to convenient coordinates near the critical points of the complex structure moduli space: the orbifold, conifold and large complex structure points. The other coordinates system is defined near a regular point in the CS moduli space. This system is convenient to study the periods close to the boundaries of convergence from the critical points patches. We describe the power series and logarithmic solutions in each of the patches and the method to obtain the transition matrices to the integral symplectic basis (\ref{Pi1lcs}). This moduli space has been studied previously in  \cite{Huang:2006hq,Candelas:1990qd}. In \cite{Huang:2006hq} the periods near the conifold were obtained. Here we are interested in having the period series up to an arbitrary order in all different patches, and for this we also compute the  transition matrices between all of those patches.

Let us start by  looking at  the PF equations on the vicinity of the LCS point.  A change of coordinates from $\psi$ to $\psi^{-5}$ in $W_{\psi}/\mathbb{Z}_5^3$ was used in \cite{Candelas:1990qd}. Here we instead use the variable $z_M=\psi^{-5}5^{-5}$. We label the variable with a subindex $M$  because the LCS ($\psi=\infty$ i.e. $z_M=0$) is a point of  MUM. 
Using this variable, the PF equation takes the form 
\begin{equation}
(\theta_{M}^4-z_M(\theta_{M}+a_1)(\theta_{M}+a_2)(\theta_{M}+a_3)(\theta_{M}+a_4)) \pi_{M,i}=0,\, i=1,2,3,4, \label{PF1}
\end{equation}
with $\pi_{M,i}$ the solutions on the LCS basis, $\theta_M=z_M\partial_{z_M}$ and $a_k=k/5,\,k=1,2,3,4$.

The next change of variables we do is  $z_O=1/z_M$.  We denote this as the orbifold basis
since the orbifold point is located at  $z_O=0$. The PF equations in these coordinates read 
\begin{equation}
(-z_O/5^5\theta_O^4+(a_1-\theta_O)(a_2-\theta_O)(a_3-\theta_O)(a_4-\theta_O)) \pi_{O,i}=0,\, i=1,2,3,4,\label{PFo}
\end{equation}
with $\theta_O=z_O \partial_{z_O}$ and $ \pi_{O,i}$ the solutions in the orbifold basis.

The most relevant coordinate system  for our discussion, is defined
as $z_C=(1-z_M 5^5)$  denoting the conifold basis with the conifold 
singularity located at $z_C=0$.  
Making the change of variables in (\ref{PF1}) we obtain the PF equations
\begin{equation}
(\theta_C^4-(1-z_C)(a_1-\theta_C)(a_2-\theta_C)(a_3-\theta_C)(a_4-\theta_C))\pi_{C,i}=0,\, i=1,2,34,\label{PFc}
\end{equation}
where $\pi_{C,i}$ are the solutions in the conifold basis and $\theta_C=(z_C-1)\partial_{z_C}$. 
The three different coordinates are related to each other
via $z_M=1/z_O=5^{-5}(1-z_C)$. In the  coordinates described above around the conifold,
LCS and orbifold points, the convergence radii  of the period series solution are $1$, $5^{-5}$ and $5^5$ respectively.
\begin{figure}
\begin{center}
\includegraphics[width=.5\textwidth]{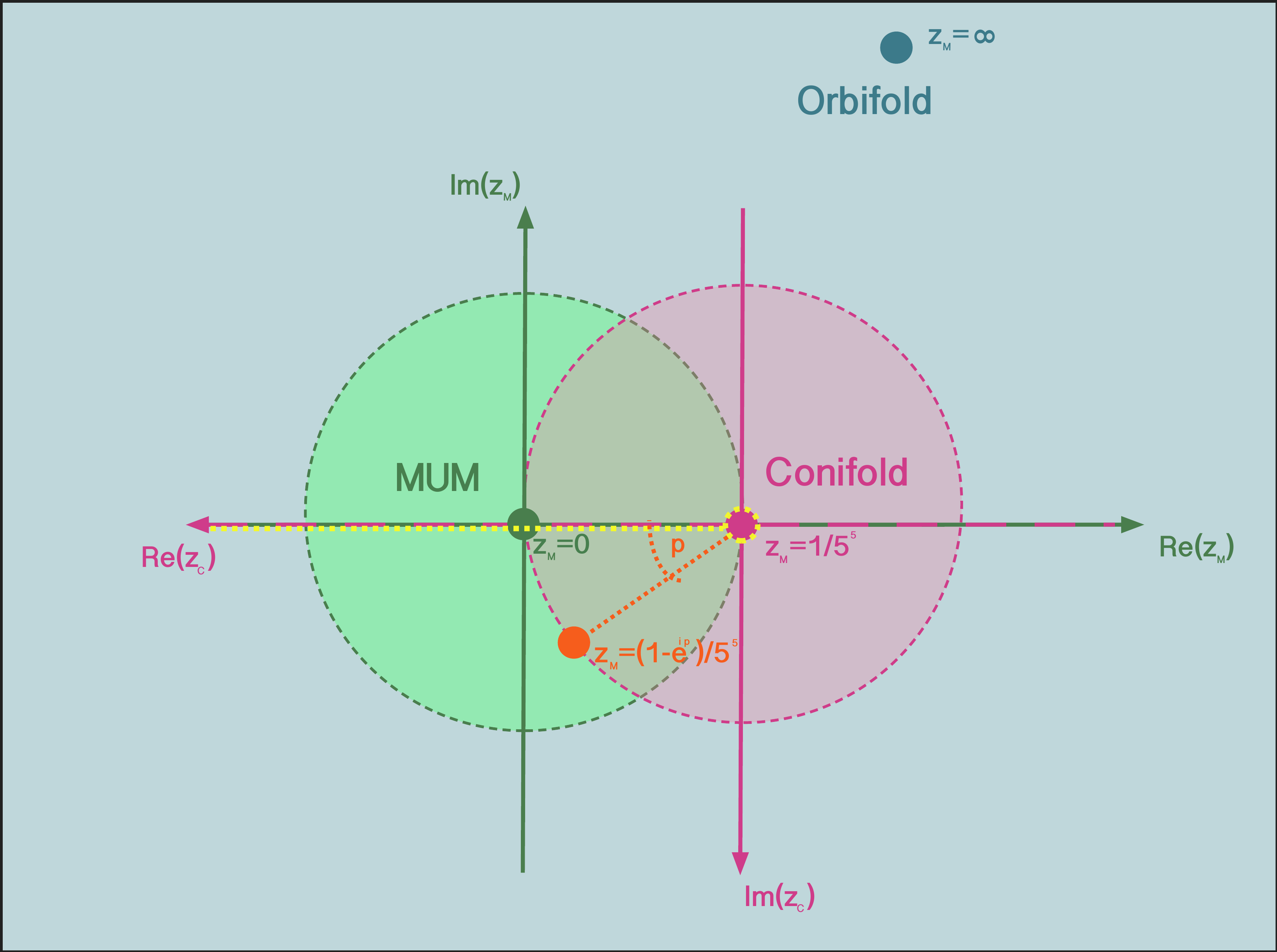}
\end{center}
\caption{The figure represents the three different critical points of the CS moduli space of the mirror of the quintic CY on $\mathbb{P}^4$
on the complex $z_M$ plane. As well we represent the regular point where we also constructed the solution to the PF equation in order to improve convergence,
$z_M=(1-e^{i p})/5^5$ with $-\pi/3<p<\pi/3$. The yellow dashed line represents the branch cut chosen for the conifold periods. Recall that $z_C=1-5^5 z_M$ and $z_O=1/z_M$. The LCS, conifold and orbifold series convergence regions are
coloured in green, pink and blue respectively.}
\label{Points1}
\end{figure}
Let us also write  the PF equations in the vicinity of an arbitrary point lying on the boundary of the conifold
convergence region. That is, we write the conifold coordinates of that point as $z^0_C=e^{i\alpha}$.
These new coordinates allow us to study the potential with precision
close to the limit of convergence of the conifold coordinates. 
We represent the critical points as well as this regular point in Figure \ref{Points1}.

The coordinates of a given point in the conifold and $\alpha$ coordinates are related as $z_C=z_\alpha+e^{i\alpha}$. Making this change of variable in (\ref{PFc}) one obtains the PF equations in terms of  the $z_{\alpha}$ coordinates:
\begin{equation}
(\theta_\alpha^4-(1-z_\alpha-e^{i\alpha})(a_1-\theta_\alpha)(a_2-\theta_\alpha)(a_3-\theta_\alpha)(a_4-\theta_\alpha))\pi^\alpha_i=0,\, i=1,2,3,4,
\end{equation}
where $\pi^{\alpha}_i$  are the solutions in the regular point basis and $\theta_\alpha=(z_\alpha+e^{i\alpha}-1)\partial_{z_\alpha}$. The point $\alpha=\pi$
constitutes the LCS point. The convergence radius of the series power solutions around $e^{i\alpha}$ is 1.
We solve the PF equations in the vicinity of the points $e^{i\alpha}$. For 
this set of points, there is  convergence of both the LCS and conifold coordinate series
when $-\pi/3<\alpha<\pi/3$. This reads
\begin{eqnarray}
z_\alpha=e^{i\alpha}-e^{i\pi/3},\qquad  z_M=1-e^{i\alpha},\qquad  z_C=e^{i\alpha},  \qquad -\pi/3<\alpha<\pi/3\,.\nn
\end{eqnarray}
This extra coordinate system serves in the study of vacua which are very close to the conifold
or LCS convergence regions. We use them to discard fake solutions appearing as a consequence
of cutting the series of the periods without achieving convergence. 

\subsubsection*{Solving the PF equations}

The solutions to the PF equations  are generalised hypergeometric
functions \cite{Candelas:1990qd,Greene:1989cf} and have been previously studied in the literature in different bases.
In \cite{Candelas:1990qd} the authors  studied the periods in the LCS and orbifold convergence regions and gave an analytic expression valid for both regions. In \cite{Huang:2006hq} the authors computed the periods on the conifold convergence region
and the transformation matrix to the integral symplectic basis, they explicitly gave the series to order five.
Here we determine the solutions near all the critical points and in particular near the conifold point, up to order 600.
This allows us to achieve  convergence in our calculations when looking for vacua and inflationary regions. 
The transition matrices that we compute in this way, match those in \cite{Huang:2006hq}.  
In the following we describe how to obtain the solutions of the PF
equations in the bases $z_M, z_C, z_O$ described previously, and the Ans\"atze employed.

Through this section we denote the independent solutions of the PF equations 
by $\pi$, while recall that  $\Pi$  denotes the periods in the integral symplectic basis (\ref{Pi1lcs}). First, we search for solutions near the conifold point, making the power series Ansatz 
\be
\pi_{C,i} = z_C^x(c_0+c_1 z_C +c_2 z_C^2+ \dots + c_n z_C^n),\, \, i=1,2,3.\label{ansatz1}
\ee
Applying the PF operator (\ref{PFc}) to (\ref{ansatz1}), the solutions for the initial equation are $x=0,1^2,2$. 
The degeneracy of  $x$ at $x=1$ indicates the existence of a logarithmic
solution. This solution vanishes when $z_C \to 0$ and can be constructed as 
\be
\pi_{C,4} = z_C(c_0+c_1 z_C +c_2 z_C^2+ \dots + c_n z^n) \ln z_C+ z_C^{x_b} (b_0+b_1 z_C +b_2 z_C^2+ \dots ).\label{powerA}
\ee

Substituting (\ref{ansatz1}) into (\ref{PF1}) we  obtain a set of recursive equations for the coefficients to each order in the expansion.  For example, in the case $x=0$ the first two equations are
\begin{eqnarray}
0&=&\frac{19\, c_0}{5}-\frac{74 \,c_1}{5}+12 \,c_2,\nn \\
0&=& -\frac{5399 \,c_0}{625}+66 \,c_1-\frac{642\, c_2}{5}+72\, c_3.\nn\\
\end{eqnarray}
The system of  equations is solved recursively. To order six,  we obtain the expressions 
\begin{eqnarray}
\pi_{C,1}&=&1 + \frac{2}{5^4} z_C^3 + \frac{97}{2\cdot 3\cdot 5^4} z_C^4 + \frac{2971}{2\cdot 3\cdot 5^5} z_C^5 + \frac{13\cdot 1175173}{2\cdot 3\cdot 5^{11}\cdot 7}z_C^6+O(z_C^7),\label{Pcon}\\
\pi_{C,2}&=&z_C + \frac{7}{10} z^2 +\frac{41}{75} z_C^3 + \frac{1133}{4\cdot 5^4} z_C^4 + \frac{6089}{5^6} z_C^5+\frac{7\cdot 13\cdot 29\cdot 61}{2\cdot 3\cdot 5^7}z_C^6+O(z_C^7), \nn \\
\pi_{C,3}&=&z_C^2 + \frac{37}{30} z_C^3 + \frac{2309}{1800} z_C^4 + \frac{31\cdot 9241} {2^3 3^2 5^5}z_C^5+\frac{41932661}{2^4\cdot 3^3\cdot 5^7}z_C^6+O(z_C^7), \nn \\
\pi_{C,4}&=&-\frac{23}{360} z_C^3 - \frac{6397}{3\cdot 10^6} z_C^4 -\frac{333323 }{2^5\cdot 5^7}z_C^5+\frac{103\cdot 353\cdot 929}{2^5\cdot 5^7}z_C^6 + \pi_{C,2} \ln z_C +O(z_C^7).\nn
\end{eqnarray}
To achieve convergence in our calculations, we use the solutions (\ref{Pcon}) up to order  600 in our analysis. The convergence radius of the power series is obtained from the formula $\lim_{n\to \infty}|c_n/c_{n+1}| =1$  and similarly for $b_n$, $\lim_{n\to \infty}|b_n/b_{n+1}| =1$.  The expressions (\ref{Pcon}) for the periods near the conifold are particularly useful for our study\footnote{In \cite{Candelas:1990qd} an analytic
integral expression for the periods valid in the orbifold and LCS convergence region was given.
For our exploration this expression is not enough, because we are interested in looking
at the behaviour of the potential near the conifold. }.

Let us now study the solutions of the PF equations near the orbifold (\ref{PFo}). This set of solutions is also obtained
by starting  with an ansatz $\pi_{O,i}=z_O^x(c_0+c_1 z_O+c_2 z_O^2+...)$ and applying to it the PF operator (\ref{PFo}). The
terms multiplying each power of $z$ need to vanish. From the $z^0$ term one gets $\prod_{i=1}^4(x-i/5)=0$,
giving the $x$ solutions $x=\frac{1}{5},\frac{2}{5},\frac{3}{5}$ and $\frac{4}{5}$. For every value of $x$ we plug the ansatz
back into (\ref{PFo}) and make some coefficients choice, to obtain the solutions
\begin{eqnarray}
\pi_{O,1}&=&z_O^{1/5} + \frac{1}{2^3\cdot3\cdot5^6} z_O^{6/5} + \frac{3}{2^3\cdot5^{12}\cdot7} z_O^{11/5}+O(z_O^{16/5}),\label{Porb}\\
\pi_{O,2}&=&z_O^{2/5} + \frac{2}{3^2\cdot 5^6} z_O^{7/5} + \frac{2401}{2^3\cdot 3^4 \cdot 5^{12}\cdot 11} z_O^{12/5}+O(z_O^{17/5}),\nn \\
\pi_{O,3}&=&z_O^{3/5} + \frac{27}{2^3\cdot 5^6\cdot 7} z_O^{8/5} + \frac{64}{5^12\cdot 7\cdot 11} z_O^{13/5}+O(z_O^{18/5}),\nn \\
\pi_{O,4}&=&z_O^{4/5} + \frac{16}{2^2\cdot 13\cdot 631}z_O^{9/5} + \frac{1458}{5^{12}\cdot 7\cdot 11\cdot 13}z_O^{14/5}+O(z_O^{19/5}).\nn
\end{eqnarray}

We  proceed in a similar fashion to  find the solutions near the LCS point in the $z_M$ variables. That is, we start by making a power series Ansatz  $\pi_{M,1}=z^x(c_0+c_1 z+c_2 z^2+...)$ plug it into (\ref{PF1}) and find the solutions  $\pi_{M,1}$.  As before, the degeneracy of $x$ in the solutions to the initial equation $x^4=0$, coming from the $z^0$ power, indicates the presence of logarithmic solutions.  One continues then with three additional Ans\"atze $\pi_{M,2},\pi_{M,3} $ and $\pi_{M,4}$, such that all $\pi_M$  have a polynomial term $w^0_M, w^1_M, w^2_M$ and $w^3_M$.  The solutions  to equation (\ref{PF1}) in the LCS point vicinity are then given by 
\begin{eqnarray}
\pi_{M,1}&=& w^0_M,\label{Plcs}\\
\pi_{M,2}&=&w^1_M +w^0_{M}\ln z_M, \,\nn\\
\pi_{M,3}&=&\frac{5}{2} w^2_M+\frac{5}{2}w^0_{M}(\ln z_M)^2+ 5 w^1_{M}\ln{z_M} ,\,\nn\\
\pi_{M,4}&=&\frac{5}{6}w^3_M+\frac{5}{6} w^0_M (\ln z_M)^3+\frac{5}{2} w^1_{M} (\ln z_M)^2+\frac{5}{2} w^2_{M}\ln z_M ,\,\nn
\eea
where the power series in them are obtained to be
\bea
w^0_M&=&\pi_{M,1}= 1 + 120 z_M + 113400 z_M^2+O(z_M^3), \nn \\
w^1_M&=&770 \,z_M + 810225 \,z_M^2+O(z_M^3),\nn\\
w^2_M&=&2875\, z_M + \frac{21040875 \,z_M^2}{4}+O(z_M^3),\nn\\
w^3_M&=& -5750 \,z_M - \frac{16491875\, z_M^2}{4}+O(z_M^3).\nn
\end{eqnarray}
From (\ref{Plcs}) one can write the explicit form of the monodromy in this basis\footnote{
The LCS monodromy in the basis (\ref{Plcs}) is given by \[
\left(
\begin{array}{cccc}
1&0&0&0\\
2\pi i&1&0&0\\
-10\pi^2&10\pi i&1&0\\ 
-\frac{20 i\pi^3}{3}&-10\pi^2&2\pi i& 1  
\end{array}
\right).
\]}. In this way it is possible to compute the periods fully  in the vicinity of all of the singular points, $\psi =0,1,\infty$, which are $z_O=0$, $z_C=0$ and $z_M=0$. We obtain the transition matrices to connect the three convergence regions
by taking sample points that lay at the intersection of the convergence regions
of the orbifold-conifold and conifold-LCS and orbifold-LCS. These are shown in Appendix \ref{transition1}. 
Changes of variables from (\ref{Plcs}), (\ref{Porb}) and (\ref{Pcon}) are made to express the periods in the integral symplectic basis 
of \cite{Candelas:1990rm} given in (\ref{Pi1lcs}). The periods in this integral symplectic basis for the three coordinates
patches $\Pi_C, \Pi_M$ and $\Pi_O$ are given in formulae (\ref{PiC}) (\ref{PiM}) and (\ref{PiO}) respectively. These explicit formulae in the three different variables allow us to determine the periods, and therefore the scalar potential,  in the full CS moduli space up to an arbitrary order.

\begin{figure}
\begin{center}
\includegraphics[width=.6\textwidth]{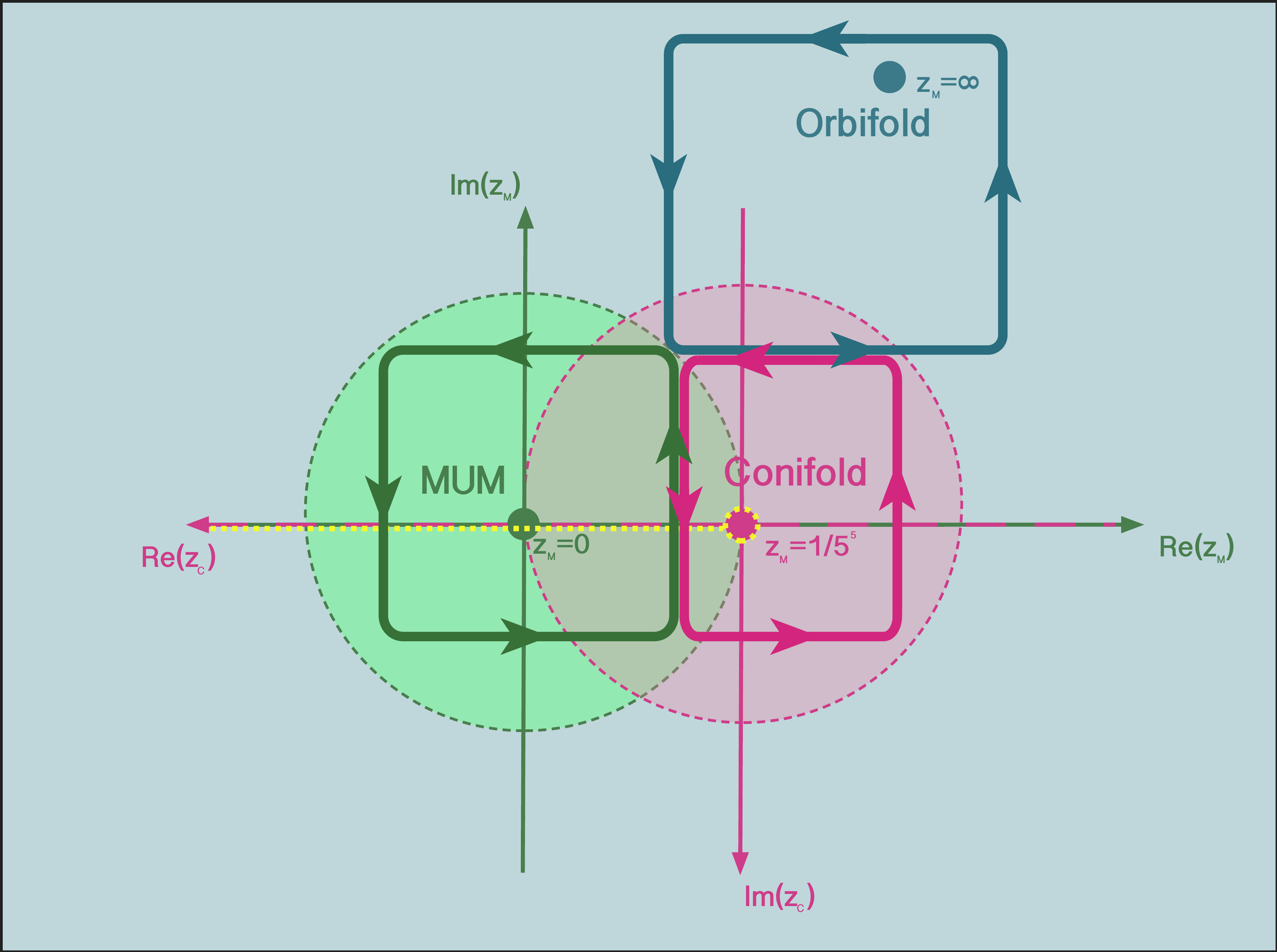}
\end{center}
\caption{The figure represents three paths leading to monodromies around the critical points of the CS moduli space of the mirror of the quintic CY on $\mathbb{P}^4$
on the complex $z_M$-plane. The paths around the LCS, conifold and orbifold points leading to monodromies $\mu_M$, $\mu_C$ and $\mu_O$ are green, pink and blue respectively. }
\label{monodromias}
\end{figure}

In the integral symplectic basis the period near the conifold $\Pi_{C,3}$ can be expressed as 
\begin{equation}
\Pi_{C,3}=-\frac{1}{2\pi i}\,\Pi_{C,1}\ln z_C+ Q(z_C),\label{mon1}
\end{equation}
where $Q(z)$ is a power series in $z$.  This can be seen from (\ref{PiC}). From (\ref{mon1}) one reads the monodromy around the conifold which is given by
\begin{equation}
\mu_C=
\left(
\begin{array}{cccc}
  1& 0  & 0  & 0\\
  0&  1 & 0  & 0\\
  -1&  0 & 1  & 0\\
  0&  0 & 0  & 1
\end{array}
\right).\label{Mu1}
\end{equation}
The periods in the integral symplectic basis in the LCS- and orbifold convergence regions are given to first orders 
in (\ref{PiM}) and (\ref{PiO}).  From these expressions one can compute the monodromy matrices around the LCS- and the orbifold point on this basis as
\begin{eqnarray}
\mu_{M}=\left(
\begin{array}{cccc}
1&-1&5&3 \\
0&1&-8&5\\
0&0&1&0\\
0&0&1&1  
\end{array}
\right), \qquad  
\mu_{O}=\left(
\begin{array}{cccc}
1&-1&5&3 \\
0&1&-8&5\\
-1&1&-4&3\\
0&0&1&1  
\end{array}
\right).\label{MonMo}
\end{eqnarray}
The relation between monodromies around the three critical points is given by $\mu_C\cdot \mu_M \cdot \mu^{-1}_O=\mathbbm{1}$.
In Figure \ref{monodromias} we represent three different paths in CS moduli space giving rise to conifold, LCS and orbifold monodromies.

\subsection{Symmetries of the potential}
\label{symmetries}

In this section we review the symmetries of the K\"ahler potential due to transformations of the moduli and how these  are broken by the superpotential generated by the fluxes. 

First of all, there is a shift symmetry in the real part of the axio-dilation, the 0-form $C_0 \to C_0 + b$, which is part of the 
$SL(2,\mathbb Z)$ symmetry of the theory (see e.g.~\cite{polchis}). Under this shift symmetry, the 3-form flux $G_{(3)}$ remains invariant, which requires $F_{(3)}$ to transform. Therefore, by keeping the fluxes fixed and transforming the axio-dilaton, the shift symmetry is spontaneously broken.

Similarly, there is a shift symmetry in the phase of the complex structure when going around the conifold,  $\theta \to \theta + 2\pi n $ ($z=r e^{i\theta}$),   with $n \in\mathbf{Z}$. This is a monodromy shift
given by $n$ powers of $\mu_C$ in (\ref{Mu1}) under which the period $\Pi_3$ transforms as
\begin{equation}
\Pi_3\rightarrow \Pi_3-n\,\Pi_1,
\label{pi3}
\end{equation} 
while the K\"ahler potential (\ref{K}) remains invariant since  $\mu_C^T \Sigma \mu_C=\Sigma$. On the other hand, it is easy to check that the superpotential transforms as 
\begin{equation}\label{Wn}
W\rightarrow W- n\,G_1\Pi_1.
\end{equation}
If we also transform the fluxes as (recall that the subindices here denote the component of the flux vector, \eqref{flujos})
\begin{equation}
G_3\rightarrow G_3 - n\, G_1,
\end{equation}
 the superpotential remains invariant. Therefore,  by keeping the  fluxes fixed, this shift symmetry is spontaneously broken. 
This is interesting for cosmological applications since   a common strategy in the literature in order to find inflation in supergravity and field theory is to consider mildly breaking  a  symmetry. 
Indeed, from the discussion above, it is  natural to think that either $\theta$ or $C_0$ or a linear combination of these fields, is a good inflaton candidate, which can give rise to either natural  or power law types of  inflation. This has been the reasoning followed in \cite{Blumenhagen:2014nba,Blumenhagen:2015jva,Hebecker:2015rya,Tatsuo,Garcia-Etxebarria:2014wla}. 

Monodromies arising from surrounding the LCS and orbifold point in the CS moduli space, given by (\ref{MonMo}), are also symmetries of the K\"ahler potential. This is because the
monodromies keep the symplectic properties of the period basis i.e.~$\mu_{M,O}^T \Sigma \mu_{M,O}=\Sigma$. Again, the superpotential transforms  under $n$ mondromies around the  LCS or the orbifold,  as $W\rightarrow G \Sigma \mu_{M,O}^n\Pi$. This implies that the fluxes have to transform as $G \rightarrow
G (\mu_{M,O}^{-1,T})^n$ in order to keep the superpotential invariant. Keeping the fluxes fixed, the  symmetries arising from monodromies are spontaneously broken. 

Note that since the fluxes are integers, the symmetry breaking will in general be hard to fine tune, unless one also fine tunes  the vevs of the other moduli. We will come back to the points discussed here in Section \ref{inflation}.




\section{Mirror quintic flux vacua:  hierarchies and inflation}\label{HI}
\label{mirrorQ}

We have now all ingredients to study the structure of  no-scale vacua of the mirror quintic in $\mathbb{P}^4$ and their  potential applications for inflation. 
We start by revisiting the vacua near the  conifold point  previously studied in the literature,
giving hierarchies among the spacetime- and compactification physical scales in the warped geometry. 
We  then describe our search for vacua using the  exact periods and describe their properties and differences with respect to vacua found using only the leading term in the series expansion. 
We finalise this section with the exploration of slow-roll inflation  in this model.

As we saw in Section \ref{sugra}, the ${\cal N}=1$ supergravity K\"ahler potential and superpotential for the dilaton and the complex structure modulus of the mirror quintic are given by 
\be\label{KW}
K = -\ln \left[-i (\tau -\bar\tau)\right] - \ln\left[-i\, \bar \Pi^T \,\Sigma \,\Pi \right] \,, \qquad  \qquad W = G \,\Sigma\, \Pi  \,,
\ee
with $G=F-\tau H $,  so that the scalar potential is 
\be\label{V1}
V =\frac{1}{2\kappa_{10}^2 g_s}\,e^K\left[K^{i\bar j}D_i W \overline{D}_{\bar j} \overline{W} \right]\,,
\ee
with $i= \tau, z$ and recall that $\Pi(z)$ are the periods given in the Appendix \ref{transition1}. 
We  now look for  non-supersymmetric  Minkowski vacua  of \eqref{V1} with  $D_\tau W=D_zW=0$  and $W\neq 0$ . The condition $D_\tau W=0$ gives an expression for $\tau$ in terms of the complex structure modulus as 
\be\label{tT}
\tau_\tau(z) = \frac{F\Sigma\bar \Pi}{H\Sigma\bar \Pi}\,, \ee
 fixing $\tau$ as a function of $z$.  Alternatively, one can find $\tau(z)$ from the condition $D_{z}W=0$: 
\be\label{tZ}
\tau_z(z) = \frac{F\Sigma(\widetilde{\Pi}\Sigma\bar \Pi)}{H\Sigma(\widetilde{\Pi}\Sigma\bar \Pi)},
\ee
with $\widetilde \Pi=\partial_z \Pi  \otimes \Pi-\Pi  \otimes \partial_z \Pi$ (observe that $\widetilde \Pi$ has two indices). The subindices $\tau$ and $z$ in (\ref{tT}) and (\ref{tZ}) indicate that $\tau_{\tau}(z)$  and $\tau_{z}(z)$ are the axio- dilaton profiles obtained from $D_{\tau} W=0$ and $D_zW=0$ respectively. Equating both expressions $\tau_z=\tau_\tau$ fixes  the complex structure $z=z_0$,  for which the fluxes  are constrained to satisfy the well-known condition $\ast G_3=- iG_3$. When there is  a solution, this is located at $\tau(z_0)=\tau_0$ and $z=z_0$, and it is a  Minkowski non-supersymmetric vacuum\footnote{Recall again that the K\"ahler moduli are at this point not stabilised, but we are interested in the dynamics of the dilation and complex structure.} with $V=0$ provided  $W\neq 0$.

A   strategy to find Minkowski vacua consists of setting the real and imaginary parts  of $\tau_\tau-\tau_z$  to zero. For our numerical analysis, it is  convenient to work only with the numerator of this quantity. This allows us to avoid fractional values of polynomials in $z$ and $\ln z$ appearing in $D_z W$ and $D_\tau W$.
The numerator of $\tau_\tau-\tau_z$ is given by
\begin{align}
A_0&=(H\Sigma\bar\Pi)(F\Sigma(\widetilde\Pi\Sigma \bar\Pi))-(F\Sigma\bar\Pi)(H\Sigma(\widetilde\Pi\Sigma \bar\Pi)).\label{ecu}
\end{align}
With the purpose of finding different Minkowski vacua configurations, we also perform 
conifold monodromies on the complex structure phase, for which the K\"ahler potential is invariant,  but as  already mentioned, this symmetry is broken by the superpotential due to the fluxes. Specifically the period $\Pi_3$ in the conifold basis transforms under a conifold monodromy as in (\ref{pi3}), 
and since the superpotential transforms as in \eqref{Wn}, the scalar potential changes as well.
Therefore,  new vacua can be found by transforming (\ref{ecu})  under $n$ conifold monodromies changing the periods via $\Pi\rightarrow \mu_C^n\Pi$.
The new expression for the numerator of ($\tau_z-\tau_{\tau}$) reads
\begin{align}
A_n=(H\Sigma\mu_C^n\bar\Pi)(F\Sigma(\mu_C^n\widetilde\Pi\Sigma \bar\Pi))
-(F\Sigma\mu_C^n\bar\Pi)(H\Sigma(\mu_C^n\widetilde\Pi\Sigma \bar\Pi)).\label{An}
\end{align}
In the next section we use $A_n$ to search for vacua and inflationary regions of the potential.

\subsection{Hierarchies revisited}
\label{hierarchies}

 The structure of the vacua for a flux configuration where the only non-vanishing flux components are $F_1, H_3$ and $H_4$, with $H_3\gg H_4$  in a type IIB CY orientifold compactification, was found in  \cite{Giddings:2001yu}. That study was performed 
at leading order in the periods series expansion near the conifold point in the CS moduli space.  
 Here we analyse this family of solutions by taking into account higher order terms in the period series.  First, we explore modifications to the zeroth   order hierarchy expression of \cite{Giddings:2001yu}, coming from constant contributions to $D_z W$. 
 We then search for the  exact vacua using the power series of the periods up to convergence order and compare these vacua with the approximations of  \cite{Giddings:2001yu}. 
Next we review the near   conifold
vacua solution of \cite{Ahlqvist:2010ki}, and compare it with our exact vacua solutions.

The warped metric which preserves Poincar\'e symmetry reads \cite{Giddings:2001yu}
\begin{eqnarray}\label{metric}
ds^2_{10} = e^{2A(y)}\eta_{\mu\nu}dx_{\mu}dx_{\nu} + e^{-2A(y)}\tilde g_{mn}dy^mdy^n.
\end{eqnarray}
The hierarchy between the spacetime (4D) and compactification (6D) physical scales will be given
by the distance of the vacuum $z_0$ to the conifold as $e^{A}\sim z_0^{1/3}$ \cite{Giddings:2001yu,comments}. In the following we write for our case the solution for $z_0$ and $\tau_0$ when only the fluxes $F_1,H_3$ and $H_4$ 
are on and $H_3 \gg H_4$. The  superpotential (\ref{fluxW}) for this case in terms of $\Pi_C(z_C)$ is given by 
\begin{eqnarray}
W=F_1 \Pi_3+\tau(H_3\Pi_1+H_4\Pi_2).
\end{eqnarray}
The periods were denoted by $\Pi_i$ instead of $\Pi_{C,i}$,
in the rest of this section we adopt this simplified notation, also $z_C$ will be denoted by $z$. 
The  value of the axio-dilaton arising from the condition  $D_{\tau} W=0$ is
\begin{eqnarray}
\tau_0=-\frac{F_1\bar \Pi_3^0}{H_4 \bar \Pi_2^0},\label{tGPK}
\end{eqnarray}
where $\Pi_i^0=\lim_{z\to 0}\Pi_i$ and $\partial_z{\Pi}_i^0=\lim_{z\to 0}\partial_z{\Pi}_i$ and  $z=0$ is the conifold point.
The third period component in the conifold convergence region, given in (\ref{PiC}), reads $\Pi_3=\Pi^0_3+\alpha z+\beta z \ln z+O(z)$. 
To estimate  the complex structure value at the minimum, $z_0$, we consider the leading terms in $D_z W=0$. 
Let us first note that the derivative of $\partial_z W$ evaluated at (\ref{tGPK}) is given by
\begin{eqnarray}
\partial_z W|_{z=z_0}&=& F_1 \, \partial_z{\Pi}_3-\frac{F_1\bar \Pi_3^0}{H_4 \bar \Pi_2^0}
(H_3 \partial_z \Pi_1^0+H_4\partial_z\Pi_2^0),\\
&=&F_1(\alpha+\beta+\beta\ln z)-\frac{F_1\bar \Pi_3^0}{H_4 \bar \Pi_2^0}(H_3 \partial_z{\Pi}_1^0+H_4\partial_z{\Pi}_2^0)+O(z).\nn
\end{eqnarray}
The covariant derivative then  reads
\begin{eqnarray}
D_z W&=&\partial_z W+\partial_z K_0 W_0 +O(z),\nn\\
&=& F_1\beta\ln z+\tau_0H_3 \partial_z{\Pi}_1^0+a_0+O(z),
\eea 
where 
\bea
W_0&=&(F_1 \Pi_3^0+\tau_0(H_3\Pi^0_1+H_4 \Pi^0_2)),\nn \\
\partial_z K_0&=&-(\bar\Pi^0_2\partial_z{\Pi}^0_4-\bar\Pi^0_4\partial_z{\Pi}^0_2-\bar \Pi_3^0\partial_z{\Pi}^0_1)/(\bar\Pi^0_2\Pi^0_4-\bar\Pi^0_4\Pi^0_2),\nn \\
a_0&=&F_1(\alpha+\beta)+\tau_0 H_4\partial_z{\Pi}^0_2+\partial_z K_0 W_0. \label{a0}
\eea
We use these relations to define a parameter $\delta_0$, which will allow us to measure the departure from our result to that in \cite{Giddings:2001yu}, as:
\bea
\delta_0&=&\frac{a_0}{F_1}=\alpha+\beta-\frac{\bar\Pi^0_3}{\bar\Pi^0_2}\partial_z{\Pi}^0_2+\partial_z K_0 \Pi^0_3-\partial_z K_0\frac{\bar\Pi^3_3}{\bar \Pi^0_2}\Pi^0_2\,, \nn
\end{eqnarray}
where we used \eqref{tGPK} and $W_0$ in \eqref{a0}.  

\begin{figure}%
\begin{center}
\includegraphics[width=.8\textwidth]{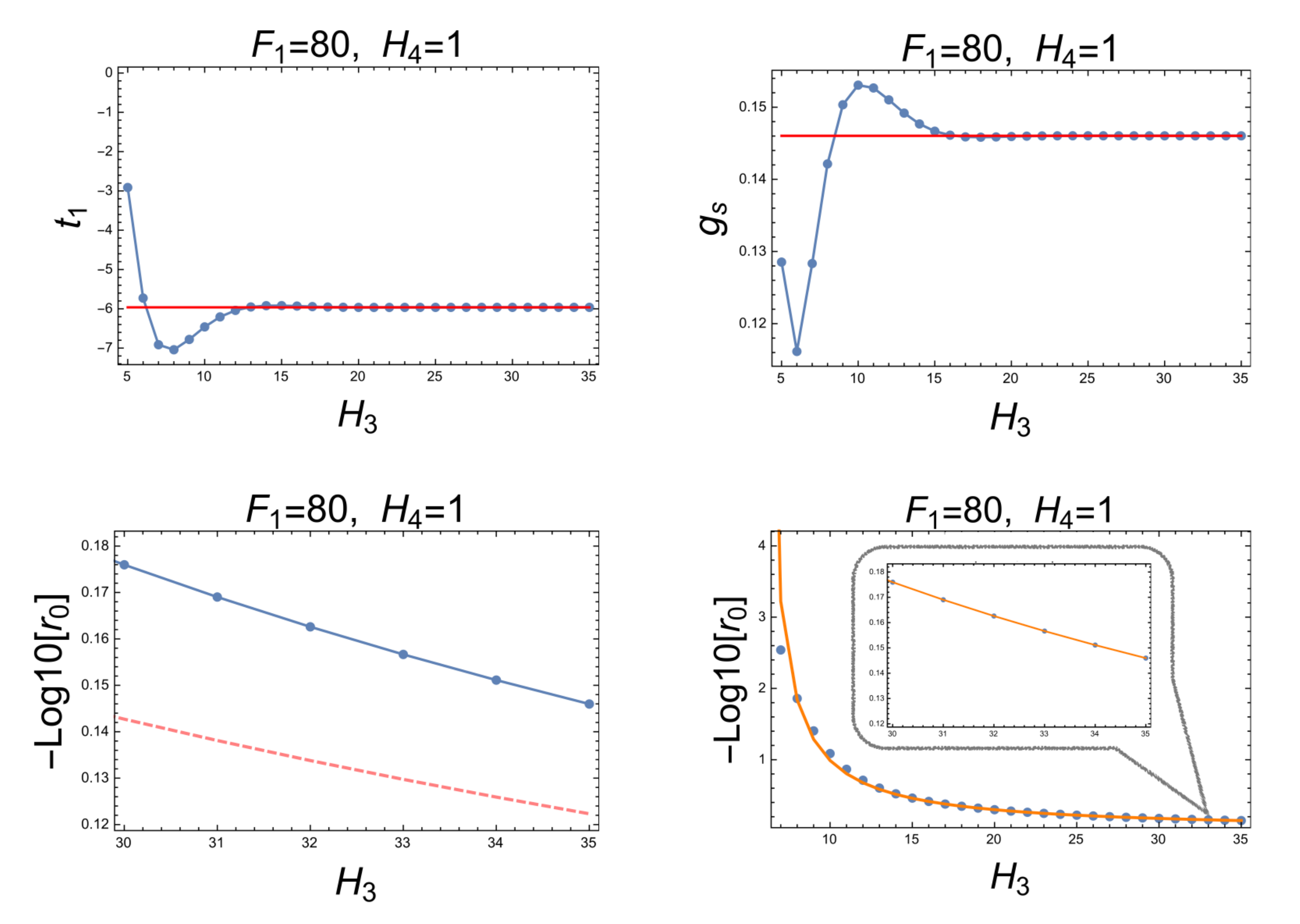}
\caption{The plots on the first row show $t_1$ and $g_s$ vacua true values (dots) for the set of fluxes $F_1=80$, $H_4=1$ and variable $H_3$, with the red line representing the hierarchical solution of \cite{Giddings:2001yu}. First plot on the second row represents the absolute value of $|z_{old}|$ (\ref{zH}) for the solution of \cite{Giddings:2001yu} (dashed line) compared with the true vacua solutions $|z_0|$ (dots).  Second plot on that row represents $|z_{new}|$, the corrected equation (\ref{zHok}) (yellow line) compared with the true vacuum $|z_0|$(dots).}
\label{conv1}
\end{center}
\end{figure}

Substituting the value of $\Pi_1$ at the conifold point, 
 $\Pi^0_1=0$, neglecting  $O(z)$  and for the moment,  $\delta_0$  terms (which are $O\left(\frac{H_3}{H_4}\right)$) one gets\footnote{The derivative of $K$ reads $\partial_z K=-\frac{\bar \Pi^T\Sigma \partial_z \Pi}{\bar \Pi^T \Sigma \Pi}$, and closed to the conifold its more relevant contribution would come from $\bar \Pi_1 \partial_z\Pi_3\sim \bar \Pi^0_1 \beta \ln z $, but $\bar \Pi_1^0=0$. 
Therefore the  most relevant contribution is the constant term $\partial_z K_0$.} 
\begin{eqnarray}
z_{old}=\exp \left(\frac{H_3}{H_4}\frac{\partial_z{\Pi}^0_1\bar\Pi^0_3}{\beta\bar{\Pi}^0_2}\right)=\exp \left(-\tau_0\frac{H_3}{F_1}\frac{\partial_z{\Pi}^0_1}{\beta}\right).\label{zH}
\end{eqnarray}
This is the result obtained in  \cite{Giddings:2001yu}, written in our notation. Let us now see how this result changes when we include the corrections due to the parameter $\delta_0$. 
When we take into account this correction,  \eqref{zH} becomes 
 \begin{eqnarray}
z_{new}=\exp \left(\frac{H_3}{H_4}\frac{\partial_z{\Pi}^0_1\bar\Pi^0_3}{\beta\bar{\Pi}^0_2}-\frac{\delta_0}{\beta}\right).\label{zHok}
\end{eqnarray}
Therefore we see that there is an extra factor   $\exp\left(-\frac{\delta_0}{\beta}\right)\sim 20$, due to the neglected terms contributing to $D_z W=0$. Using the  expressions above, we can now check the effects of the corrections due to $\delta_0$. 
We show this in Figure  \ref{conv1}. We compare first the full value of $\tau_0$ \eqref{tGPK} at the minimum as a function of $H_3$ (the two plots in the first line) with respect to the approximated value obtained in \cite{Giddings:2001yu}, which corresponds to the constant red line in the plots. As can be seen, $\tau_0$ converges rapidly to the approximated value at the minimum as the flux is increased. Notice also that a small perturbative value of $g_s$ depends on the smallness of the ratio $F_1/H_4$.
Next we do the same for $|z_0|$ (first plot in the second row). Here it is clear that the approximated value (dotted orange line) for $|z_0|$ does not converge to the actual value (blue continuous line) even when the  flux increases. 
In the last two pictures we plot instead the true value at the vacuum (the dots) and the $\delta_0$ corrected  value $z_{new}$ (orange continuous line). The convergence is almost instantaneous. Therefore, we see that the estimated value for $|z_0|$ is better represented by our corrected expression \eqref{zHok}. In Appendix \ref{Ahierar} we write the correction (\ref{zHok}) in the notation of \cite{Giddings:2001yu}.

\subsubsection*{More generic flux configurations}

We now give  a  condition for  general configurations of  fluxes that can be used for finding vacua with  hierarchies.
In doing this, the fluxes $F_3$ and $H_3$ play an important role, because it is possible to leave the value of  the axio-dilaton fixed, by varying these fluxes, as was done in  \cite{Giddings:2001yu} for $H_3$.  The solution for $z_0$,  with $|z_0|\ll1$ at the minimum for arbitrary fluxes was given in \cite{Ahlqvist:2010ki} and there it was also found that the density of vacua near the conifold is high. 
Here we will see how varying $F_3$ and $H_3$, one can move close to the conifold point and that the true vacua approach to $z_0$ only if the condition $|z_0|\ll1$ is satisfied.

Let us first find  the approximate value of $z_0$ for a vacuum near the conifold point. 
The covariant derivative $D_z W$ close to the conifold point to leading order reads
\begin{eqnarray}
D_z W&=&(F_1-\tau_0 H_1)\beta \ln z+(F_1-\tau_0H_1)(\alpha+\beta)+
(F_2-\tau_0 H_2)\partial_z{\Pi^0_4}\\
&&\hskip 1cm  - (F_3-\tau_0 H_3)\partial_z{\Pi^0_1}-(F_4-\tau_0 H_4)\partial_z{\Pi^0_2}
+\partial_z K_0 W_0,\nn
\eea
where 
\begin{eqnarray}
\tau_0=\frac{F\Sigma \bar\Pi^0}{H\Sigma \bar\Pi^0}\,, \qquad \qquad 
W_0= (F-\tau_0 H)\Sigma \Pi^0 \,.\label{t0G}
\end{eqnarray}
(Note that $\tau_0$ here depends on the fluxes).
This  gives as a solution for $z_0$:
\begin{eqnarray}
z_0\sim \exp\left(-\left(\frac{\alpha}{\beta}+1\right)+\frac{-(F_2-\tau_0 H_2)\partial_z{\Pi}_4^0+(F_3-\tau_0 H_3)\partial_z{\Pi^0_1}+(F_4-\tau_0 H_4)\partial_z{\Pi^0_2}-\partial_z K_0 W_0}{\beta(F_1-\tau_0 H_1)}\right)\,. \nn \\ 
\label{aproxG}
\end{eqnarray}
Now, since $\Pi^0_1=0$ the  contribution of the fluxes $F_3$ and $H_3$ in $\tau_0$ is absent  and therefore we can tune these  to achieve  a small $|z_0|$, which gives hierarchies, while preserving a stabilised perturbative $g_s<1$. 
We show this explicitly  in Figure \ref{hierG} for two configurations  of fluxes. We compare the approximate value of $z_0$ given by \eqref{aproxG} and its real value using the exact periods\footnote{Where by exact  we mean that we employ as many necessary terms in the period series so as to achieve convergence, as already mentioned.}.
The figures show that the  calculation with the period expansion and the approximation
(\ref{aproxG}) differ on a $\sim 1 \%$ for $|z_0| \sim 10^{-1}$. However even for points inside the conifold convergence region, the difference is higher, for example for $|z_0|\sim 0.5$ the difference is a $\sim 12\%$.

\begin{figure}
\begin{center}
\includegraphics[width=.8\textwidth]{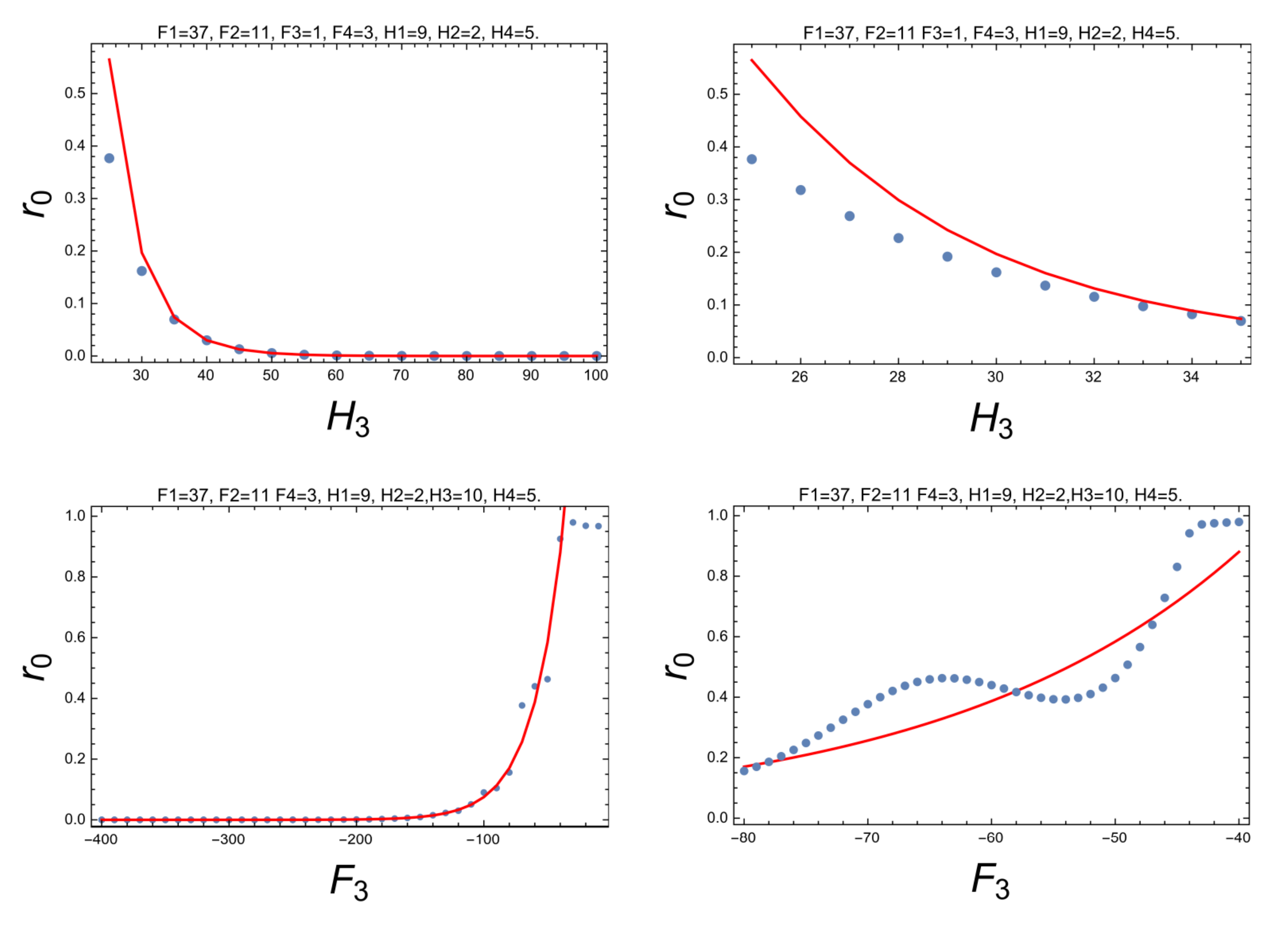}
\caption{The first plot shows the exact $|z_0|$ for vacua solutions vs.~$H_3$ (blue dots) together with approximation (\ref{aproxG}) (red line).  The second plot  is a zoom of the first,  showing more clearly the difference between the exact solution and the approximation. The third and fourth plots show the exact vacua solutions (dots)
and the approximation (\ref{aproxG}) vs.~$F_3$, and its  zoom. Turning on $H_3$ and $F_3$, leaves $g_s$ fixed and leads to vacua close to the conifold. This implies a hierarchy between the four and six dimensional scales.}
\label{hierG}
\end{center}
\end{figure}

\subsection{Search for vacua}
\label{sVacua}

We   search for Minkowski vacua for different flux configurations. Since we are looking for vacua numerically, the  solutions will depend on the approximation taken for the expansions of the periods. Through our explorations we explore  the convergence of 
$r$ up to order 600 in the expansions. 
From our search we identify different generic properties of the vacua depending on our choice of fluxes, conifold monodromies and location with respect to the critical points. These are as follows:
\begin{enumerate}
\item
We find Minkowski  vacua for which the complex structure axion value lies outside the basic domain region from 0 to 2$\pi$. We did  not find inflationary regions for models allowing these vacua. See Section~\ref{inflation}.
\item
We find {\it Fake} Minkowski  vacua, for which all moduli vevs depend on the order  of the approximation on the period series. 
\item
Inflationary regions are present for small and ``large" values
of the distance to the conifold  $r=|z_0|$. This was checked starting with  a given order in the series expansion,
and holds for an arbitrary order. No Minkowski vacua were found for the choices of fluxes in these cases. 
In one  case we found a de Sitter vacuum. We discuss this in Section \ref{inflation}.
\end{enumerate}

In all solutions, the values of the  fluxes can be tuned to achieve a perturbative value of $g_s$, $g_s<1$. We now discuss these cases in more detail. \\

\noindent
{\em I. Minkowski vacua}\\
We present here the  flux configurations for which there are stable Minkowski vacua  in a region close to the conifold point ($r<1$). Some generic features about these solutions are the following:
\begin{itemize}

\item
We found several Minkowski exact vacua within the conifold convergence region using the period series expansion $\Pi_C(z_C)$ in (\ref{PiC}).  The order in the period's series expansion is increased up to achieve convergence. In some cases we stopped at order 200, but we have obtained the period series up to order 600. In Table \ref{minC2} we present  14 of these vacua. 
%
%
%
%
\begin{table}[htdp]
\begin{center}
\begin{tabular}{|c|c|c|c|c|c|c|c|c|c|} \hline
&$r$ & $\theta$& $t_1$ & $t_2$&\footnotesize$(F_1,H_1)$&\footnotesize$(F_2,H_2)$&\footnotesize$(F_3,H_3)$&\footnotesize $(F_4,H_4)$\\ \hline
1&0.00387722& -7.01112& -2.965416&3.421883&(40,0)&(0,0)&(0,16)&(0,1)\\ \hline
2&0.289795& -3.90606&-7.0416876&7.0353577&(80,0)&(0,0)&(0,8)&(0,1)\\ \hline
3&0.289795& -3.90606&-176.04219&175.88394&(2000,0)&(0,0)&(0,8)&(0,1)\\ \hline
4&0.289795& -3.90606&-4.40105&4.3971&(50,0)&(0,0)&(0,8)&(0,1)\\ \hline
5&0.476018&-21.5600&-3.54466&5.02946&(9*10,1)&(0,0)&(27*10,16)&(0,2) \\ \hline
6&0.26791&-2.65769&-1.13736&2.11955&(20,0)&(0,0)&(0,8)&(0,1)\\ \hline
7&0.0038772&-7.01111&-4.44813&5.13282&(60,0)&(0,0)&(0,16)&(0,1)\\ \hline
8&0.0553517&-1.88428&-5.51566&20.8484&(200,1)&(30,1)&(2,10)&(2,1)\\ \hline
9&$2.07602\cdot10^{-6}$&-13.6039&-5.96259&6.84777&(80,0)&(0,0)&(0,30)&(0,1)\\ \hline
10&$0.160500$& $1.7234$&0.407671&0.81259&$(37, 9)$& $(11,2)$& $(1,31)$& $(3,5)$\\ \hline
11&$0.000301$&$7.2269$&-1.22438 &44.711&$(16, 2)$& $(7,7)$& $(1,-8)$& $(4,-1)$\\ \hline 
12&$6.28576\cdot 10^{-8}$&$-4.06$&$123.57$&$124.58$&$(36, 2)$& $(107,0)$& $(0,5)$& $(0,1)$\\ \hline 
13&$8.91875\cdot 10^{-7}$&$-47.91$&-4.75&1.56681&$(2, 0)$&$(4, -2)$&$(1, 3)$&$(1, 0)$\\ \hline 
14&$0.03351$&$6.28319$&$-3$&$3.71019$&$(3, -1)$&$(3, 0)$&$(1, 1)$&$(0, 0)$\\ \hline 
\end{tabular}
\end{center}
\caption{The table shows the  values of the moduli at the minima of the scalar potential $V$ in the conifold convergence region. Solutions for fluxes where $H_i$ and $F_i$ are proportional, have the same $z$ and have values of $\tau$ related as in (\ref{related}). Here $z_C=r e^{i\theta}$,  $\tau=t_1+i t_2$ and the fluxes are given in string units. }
\label{minC2}
\end{table}

\item
Solutions where the vacuum lies very close to the conifold point present a large hierarchy between the internal and the macroscopic dimensions, confirming the results of  \cite{Giddings:2001yu} up to the correction (\ref{zHok}). These were found considering only the leading contribution to the periods
and non-zero $F_1,H_3,H_4\neq 0$, $H_3 \gg H_4$. 
In Figure \ref{conv1} we compare the value of $z_0$ with the one of the convergent solution for 
different values of $H_3$ with fixed values of $F_1$ and $H_4$. An example  is the vacuum  9 in
Table \ref{minC2}, where $|z_0|\sim 2\cdot10^{-9}$. One can slightly deviate from these solutions by turning on nonzero values of $H_1$ or $F_3$.
Also the vacua on Table \ref{minC2} with more generic flux configuration possessing hierarchies
are obtained by the guideline of considering $|z_0|\ll 1$ in (\ref{aproxG}), examples of vacua with a  strong hierarchy are  $12$ and $13$ in the table.  

\item
For a specific flux configuration with only $F_1$ and $H_3$ non-zero it is straightforward to see from eq.~(\ref{ecu}) that the  solution for $z$ does not depend on the fluxes. Hence, if  there is a $z_0$ vacuum solution, this is unique for all set of fluxes $(F_1, H_3)$.  However,  studying the sample case
$H_3= 16$ and  $F_1 =1$ we find, in agreement with \cite{Giddings:2001yu}, that is not possible to find a vacuum close to the conifold point for this flux configuration, which in turn implies the absence of a solution for any flux configuration of this type. We have checked this using an expansion for the periods around the conifold up to order 200.  A simpler argument tells that a near to the conifold solution will have the approximate value of $\tau_0=-\frac{F_1 \bar \Pi_3^0}{H_3\Pi_1^0}\rightarrow \infty$, since $\Pi_1^0=0$.

\item Another feature of the solutions comes from the set of equations (\ref{tT}) and (\ref{tZ}), from which one can see that  two flux configurations $(F^{(1)}, H^{(1)})$ and $(F^{(2)}, H^{(2)})$ with components related by 
\begin{eqnarray}
\frac{F_i^{(1)}}{H_i^{(1)}}= k \frac{ F_i^{(2)}}{H_i^{(2)}},\qquad  \forall_i \quad\text{ with } k \in \mathbf{Q}. \label{related}
\end{eqnarray} 
will have stable vacua at the same value of the complex structure and a fixed dilaton value 
related by $\tau_0^{(1)}=k \tau_0^{(2)}$. 
This defines a similarity relation and therefore a characteristic class $[F, H]$ to which all the above fluxes belong to. 
In terms of the corresponding characteristic class, there is only a single vacuum.

\item 
To check the solutions found in the conifold convergence region in terms of $\Pi_C$  we explore them also on other patches
(LCS, orbifold, regular point).  This is particularly relevant for vacua that are apparently located close to the boundary of the 
conifold convergence region, using instead of $\Pi_C$ periods's expansions that converge faster to the solution. The periods expanded in  local coordinates near a critical point have the corresponding  monodromy  automatically. For example the $\Pi_C$ expansion has the term $\ln z_C$, causing the monodromy $\mu_C$ that is absent in $\Pi_O$ and $\Pi_M$. 
If the periods are expressed as local series expansion in the variables of other patches, 
the monodromy around the original critical point has to be implemented by hand.  
We find Minkowski vacua as zeroes of the numerator of $(\tau_z-\tau_{\tau})$,
which transforms under monodromies $\mu_C$ as in formula (\ref{An}). 
Vacua exploration on patches different from $z_C$ corresponds to finding zeroes of it. 
The periods written as a series of the complex structure modulus in 
the three different coordinate systems $z_C,z_M$ and $z_O$ are given in Appendix \ref{transition1}. 

\item
We found  that the monodromies play an important role in finding vacua. For certain sets of fluxes, the region 
$0\leq \theta < 2\pi$ does not  contain Minkowski vacua, and one can only satisfy the conditions $V=\partial_i V=0$ after
taking monodromies around the conifold point, starting from that domain. This can be seen  in Table \ref{minC2}. The vacua presented there are only found for a given value of $\theta$ but they vanish when taking  $\theta\rightarrow \theta + 2\pi n,\, n\in \mathbb{Z}$.
However, for certain configurations of fluxes, it is possible to find vacua at every monodromy turn
differing in the values of the other moduli \cite{Ahlqvist:2010ki}. 

\end{itemize}

For all the reported vacua, the corresponding mass matrix is positive definite ensuring  stability of the minima (up to K\"ahler moduli).\\

\noindent

{\em II. Fake Minkowski vacua}

Our study shows that in order to find true vacua and study their properties, it is necessary to include  higher order terms in the series expansion of the periods  in terms of the complex structure modulus. Staying at leading order leads to vacua which vanish when  higher order terms are taken into account.  Let us give an example of this: in Figure \ref{conv2}, we 
 plot the value of $|z_0|$ for two different sets of fluxes. The first has an apparent vacuum at leading order in the series expansion. However,   as higher orders are included, the apparent vacuum solution does not converge to a final value. Aditionaly the root approaches to the boundary of convergence. In the figure we see that  convergence is not reached even after taking 200 orders in the series;  the $n+1$ value
differs from the $n$ values by about $1 \%$. We checked in the orbifold patch for this vacuum and it is absent, which  tells us
that it is an error of the approximation. On the right picture we see a second configuration of fluxes for which the vacuum solution  at the leading order, remains practically unchanged after the 200th order. Indeed, in this case, convergence to the actual vacuum is achieved very quickly.

\begin{figure}%
\begin{center}
\includegraphics[width=.9\textwidth]{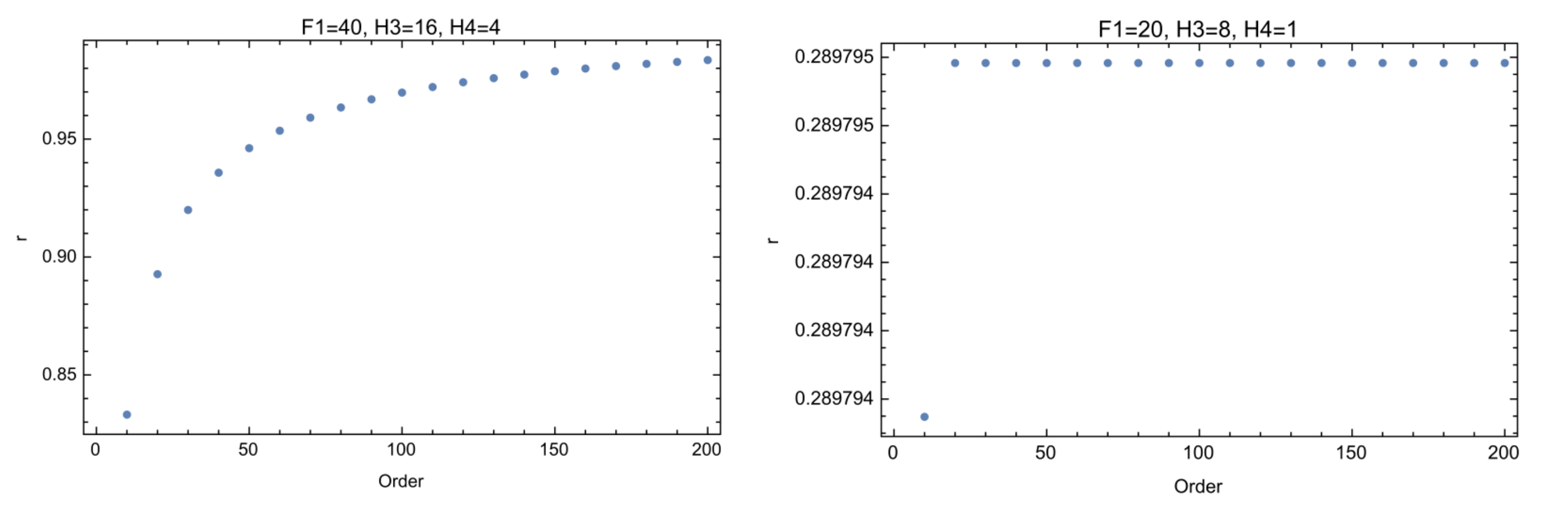}
\caption{Vacuum solutions vs. the order in the period expansion for two sets of non-zero fluxes.
The vacuum solution on the left does not converge inside the conifold convergence region after order 200th, and thus does not correspond to a true vacuum. Instead, the solution on the right  converges very quickly to a stable value.}
\label{conv2}
\end{center}
\end{figure}
%
%

\subsection{Inflationary regions}
\label{inflation}

In this section we  explore the {\em full} complex structure moduli space in order to determine whether inflationary directions can generically appear in the scalar potential. The CS moduli space has been explored in several papers in the literature. In \cite{Blumenhagen:2014nba,Hebecker:2015rya,Blumenhagen:2015jva}, axion inflation was studied in the small region of the moduli space near the LCS point at leading order in the periods' expansion, while in  \cite{Garcia-Etxebarria:2014wla} scalar potentials near the critical points were explored at leading order in the period's series expansion. 
The recent work of  \cite{Tatsuo} on the other hand, studied the scalar potential in the region close to the LCS  point, but considered only a small number of terms  in the series' expansion. We will comment further on this work below. 
A standard strategy to look for potential inflationary directions in the literature, is to track symmetries  of the K\"ahler  potential, which may be slightly broken by some effect, such as fluxes.  
In  Section \ref{symmetries}, we discussed the symmetries of the K\"ahler potential when transforming the dilaton and the phase of the complex structure moduli, in particular by a shift. 
Following this line of thought,  we first let the fields evolve and look for regions in the moduli space where the generalised slow roll conditions for multi-field inflation are satisfied mostly along the  axionic directions: the complex structure phase $\theta$ and $\Re(\tau)$. The multi-field slow-roll parameters are given by  (see e.g.~\cite{Marta1}):
\be\label{SRP}
\epsilon = M_{Pl}^2 \frac{K^{i\bar j}\nabla_i V\nabla_{\bar{j}} V}{V^2} \,, \qquad \quad \eta = {\rm min \,\,\, eigenvector} \left[ \frac{K^{i\bar j} \nabla_i\nabla_{\bar j} V}{V}\right]\,,
\ee
where $\nabla$ is the covariant derivative in the moduli space. 

Using this approach, we did not find  field regions where $\theta$ and $Re (\tau)$ have long displacements and slow-roll parameters are small.
This check was done at an arbitrary order in the series expansion of the periods. 
This can be understood in the following way. From the form of $K$ and $W$, considering all fields but $\theta$ fixed, it is easy to 
see that, due to the presence of the logarithms in the periods, the scalar potential will contain powers of $\theta$, besides sines and cosines, giving  
\be
V(\theta) \sim A + B \,\theta^2 + C\,\theta \cos \theta + \dots 
\ee
where $A, B$ depend on the other moduli and the fluxes, and the dots include further mixed terms, including sines and cosines multiplied by powers of $\theta$. This rather generic form of the potential for the complex structure axion was pointed out in \cite{Tatsuo}. However, while in \cite{Tatsuo} it is argued that this kind of potential can give rise to natural inflation, in several cases we find that the modulations of the potential along the $\theta$-direction, are too high to allow for slow roll inflation. Moreover the amplitude of the oscillations  increases with $|\theta|$ (see Figs.~\ref{mod} and \ref{mod1}).  However, as pointed out in \cite{PTZ}, it is possible that more general slow-roll regions appear in this direction, allowing for inflation.  
We also  calculated $\epsilon$  assuming single-field inflation along the $\theta$ direction, keeping all other moduli fixed. We found that in some cases it was possible to  get $\epsilon \ll1$,  however  the $r$ direction was highly unstable, and therefore the multi-field  $\epsilon$  differed very much from the single field approximation.
 This can be understood as due to  higher order corrections in the period series giving contributions to the potential
producing interaction terms between $r$ and $\theta$.

Given this result, we   explored the scalar potential in all possible directions. We searched systematically  for numerical minimisation of $\epsilon$, varying $z$, $\tau$ and the fluxes, subject to constraints on $z_0$ and $\tau_0$ on (\ref{aproxG})(\ref{t0G}), $|z_0|\ll1$ and $\Im(\tau_0)>1$ respectively. We found no flux configuration where a Minkowski vacuum near the conifold and $\epsilon  \ll 1$ occurs simultaneously. Discarding the restriction of having a vacuum near the  conifold point,  we found through numerical minimisation, small values of $\epsilon$ for certain regions of  $z,\tau$ and  flux configurations. For those cases there were no Minkowski vacua found
 on the orbifold and LCS convergence regions. We discuss below our results. 

\begin{enumerate}
\item We found that for configurations of fluxes  with a Minkowski vacuum, there are no regions where the multi-field slow roll parameters \eqref{SRP} are smaller than one.  
This is shown in Figure \ref{mod}, where there were no inflationary regions (defined as regions with $\epsilon,\eta <1$). Again, it is possible that we miss more general slow-roll regions as discussed recently in \cite{PTZ}. Indeed, the potential along the $\theta$ direction shown in Figure \ref{mod} resembles closely those discussed in \cite{PTZ}.

\begin{figure}%
\begin{center}
\includegraphics[width=.9\textwidth]{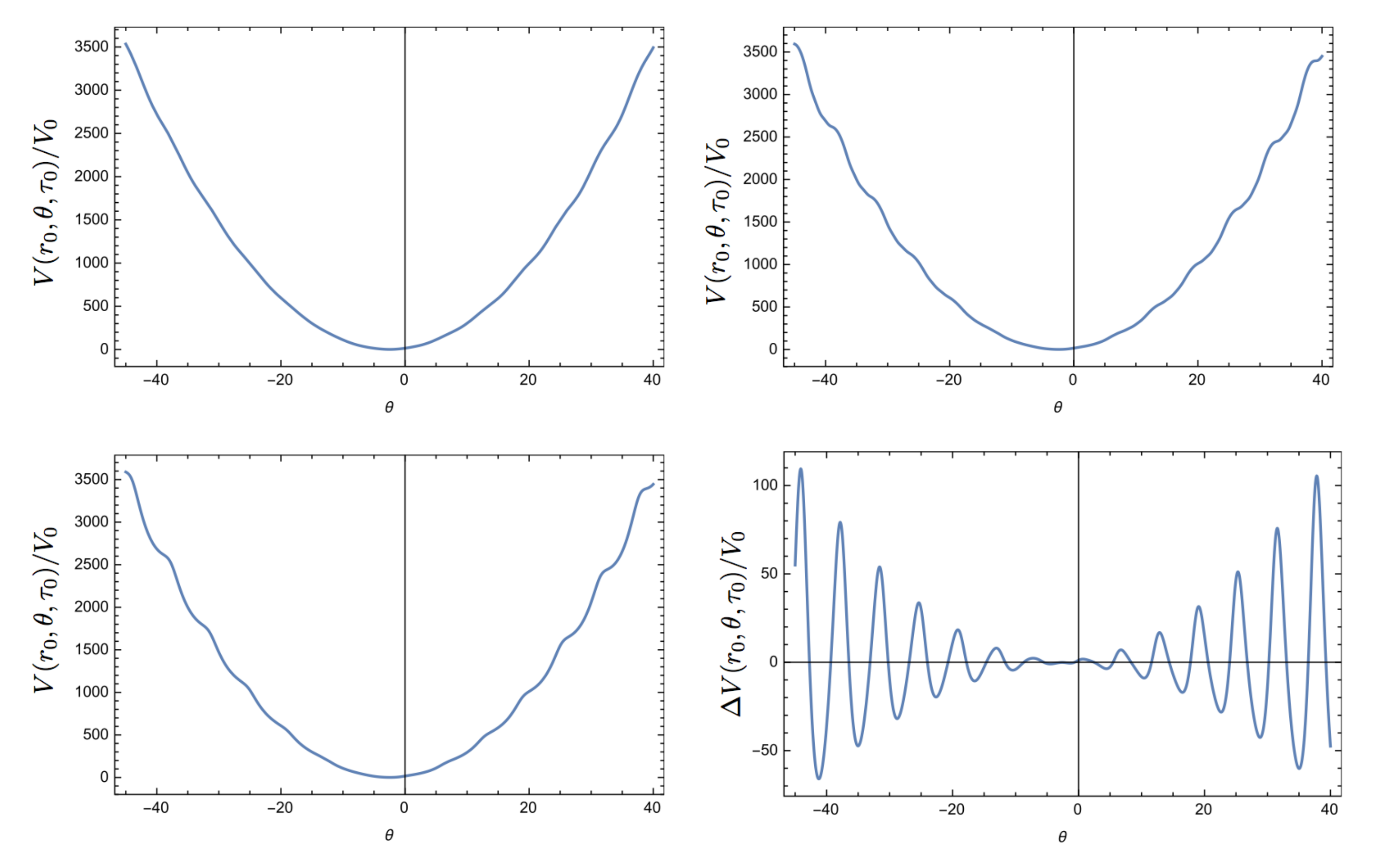}
\caption{Scalar potential vs. $\theta$ and  the rest of the moduli fixed at their vev's for a configuration with non-zero fluxes $F_1=20$, $H_3=8$, $H_4=1$ and a Minkowski vacuum. 
Here $V_0 = \alpha'^2/(2 \kappa_{10}^2 g_s)$  and the fluxes are given in string units. The values of the other moduli are set at $r_0=0.26791, t_1=-1.13736, t_2=2.11955$. The figures show the scalar potential  approximations to order 1, 4, 200 in $z$ from left to right and up to down. The last plot indicates the difference between order 200 and the order one calculation.}
\label{mod}
\end{center}
\end{figure}

\item We found configurations of fluxes (see Table (\ref{nv-inflacion}))  for which large inflationary regions are present. We show  examples of this in Figures \ref{mod-inf1}-\ref{eCon1}. 
 As in the previous case,  large modulations appear due to higher order terms in the series expansion. We show this in Figure \ref{mod1} where we also show the difference among modulations for the potential as function of $\theta$ only, when the first and fourth order terms are kept.   
In Figure \ref{mod-inf} we illustrate the effect of taking into account higher order terms in the series expansion of the periods, by plotting  the values of $\epsilon$ at different  points in the moduli space vs.~the order of the expansion. 

These configurations of fluxes however, did not give rise to Minkowski vacua since the equations $D_\tau W =D_z W =0 $, did  not have a solution. 
In the conifold convergence region this happens because the solutions to both constraints $\tau_{\tau}$ and $\tau_z$ on (\ref{tT}) and (\ref{tZ}) give imaginary parts with opposite signs.
Interestingly, for these configurations we found many saddle points with a generic feature: the main unstables direction are given by $\theta$ and $r$. 
Given this observation, we  explored inflationary regions near the orbifold singularity as we discuss below. 
The precise location of saddle points turns out to be highly affected by the order of the series expansion, but convergence is obtained. 
Additionally we found a $dS$ vacuum for a flux configuration satisfying  slow-roll conditions. This case is presented 
in Figure \ref{eCon3}.  This solution  is interesting as it may give  an explicit realisation of the uplift mechanism proposed in \cite{SS} once we include the K\"ahler moduli.

In Figures \ref{mod-inf1},  \ref{eCon3}, \ref{eCon1} and \ref{eCon2} we show examples of $\epsilon\ll 1$ values for different flux configurations on the conifold convergence region. In Table \ref{nv-inflacion} we show those flux configurations and two others where also $\epsilon\ll 1$ values were found. In all of the cases the fluxes do not satisfy the condition to encounter a hierarchical vacuum near the conifold ($|z_0|\ll 1$, $z_0$ in (\ref{aproxG})).
\smallskip

\begin{table}[htdp]
\begin{center}
\begin{tabular}{|c|c|c|c|} \hline
$F_1, H_1$& $F_2, H_2$& $F_3,H_3$& $F_4, H_4$\\ \hline
1,1&0,-10&0,1&-10,1\\ \hline
2,4&2,4&1,2&3,1\\ \hline
1,3&0,0&10,2&0,1\\ \hline
2,4&0,0&6,2&0,2\\ \hline
43,10&193,64&198,-10&-10,-10\\ \hline
90,3&193,165&-10,0&-10,0\\ \hline
\end{tabular}
\end{center}
\caption{Configurations of fluxes for which inflationary regions appear, but there are no Minkowski vacua.}
\label{nv-inflacion}
\end{table}
The slow-roll conditions found for this type of flux configurations occurred in general in a  multifield fashion.
For example in Figure~\ref{mod-inf1} the $\eta$-eigenvector along the minimum $\eta$-eigenvalue ($\eta_{min}$)
is given mostly  in the $r$ direction. On the other hand in Figure~\ref{eCon3} and Figure~\ref{eCon2} the dominant contributions
to the $\eta_{min}$ eigenvector are given by $t_1$ and $t_2$. Finally in Figure~\ref{eCon1} the dominant contribution is given
mostly along the direction of $t_1$. In  Figs.~\ref{mod-inf1}-\ref{eOrb1} we give  an estimate of the displacements
of the canonical fields $\phi_r,\phi_{\theta},\phi_{t_1},\phi_{t_2}$ 
defined as 
\begin{eqnarray}\label{canofields}
\partial_{\mu} \phi_{\theta}&=&M_{Pl}\sqrt{K_{z\bar z}} \, r\,\partial_{\mu} \theta,\  \  \partial_{\mu}\phi_r=M_{Pl}\sqrt{K_{z\bar z}}\,\partial_{\mu} r,\nn\\
\partial_{\mu} \phi_{t_1}&=&M_{Pl}\frac{\partial_{\mu} t_1}{2 t_2},\  \  \partial_{\mu}\phi_{t_2}=M_{Pl}\frac{\partial_{\mu} t_2}{2 t_2}\,,\nn
\end{eqnarray}
in the inflationary region.
We evaluate the quantities above locally\footnote{By locally here we  mean that  the values of the other moduli are frozen when  defining the  canonical field for  a  single modulus. For example for small field displacements $\delta r,\delta \theta, \delta t_1,\delta t_2$ around the point $r_0,\theta_0,t_{1,0}$ and $t_{2,0}$
the canonical fields are given by $\partial_\mu \phi_{t_1}=\partial_\mu\frac{ t_1}{2 t_2^0}+O(\delta t_1,\delta t_2)$, $\partial_\mu \phi_{t_2}=\partial_\mu\frac{ t_2}{2 t_2^0}+O(\delta t_1,\delta t_2)$
$\partial_\mu \phi_r=\partial_\mu (\sqrt{K_{z\bar z}^0}r) +O(\delta r,\delta \theta)$, $\partial_\mu \phi_{\theta}=\partial_\mu( \sqrt{K_{z\bar z}^0}r_0 \theta)+O(\delta r,\delta \theta)$. } in order to estimate the displacements of the canonical fields on the
slow-roll region. 

There seems to be no pattern indicating that slow-roll regions occurs along a preferred direction.
In particularly it doesn't occur necessarily along a direction with a shift symmetry.
For example we did not find flux configurations with slow-roll conditions where $\theta$
is the dominant slow-roll direction. This observation indicates that to achieve slow-roll along 
$\theta$ one would require a careful fine tuning of the fluxes, which will be hard to do as they are integers.

\item Finally, we also  explored slow-roll conditions on the orbifold convergence region. This exploration is motivated by our  findings  that the flat directions on the previous cases seem to extend for larger values of $r_C$, going outside the conifold convergence region with boundary at $r_C=1$.
In Figure\ref{eOrb1} we show the density  plots of $\epsilon$ on six different planes of the moduli space in the orbifold convergence region, 
for the same flux configuration as in Figure \ref{eCon2}. 
For this configuration there is a saddle point inside the orbifold convergence region.
However there are no Minkowski nor dS vacua.

Note that the  distance between critical points of the complex structure moduli space is finite. Therefore the displacements of the
canonical field for  $r$, $\phi_r$,  are bounded in $M_{Pl}$ units. We show this in Figure~\ref{Mpldisp} where we plot 
the evolution of the locally normalised  canonical  fields $\phi_r$ and $\phi_{\theta}$ vs.~the moduli $r$ and
$\theta$ for the conifold and the orbifold convergence regions.

\end{enumerate}

\begin{figure}
\begin{center}
\includegraphics[width=.9\textwidth]{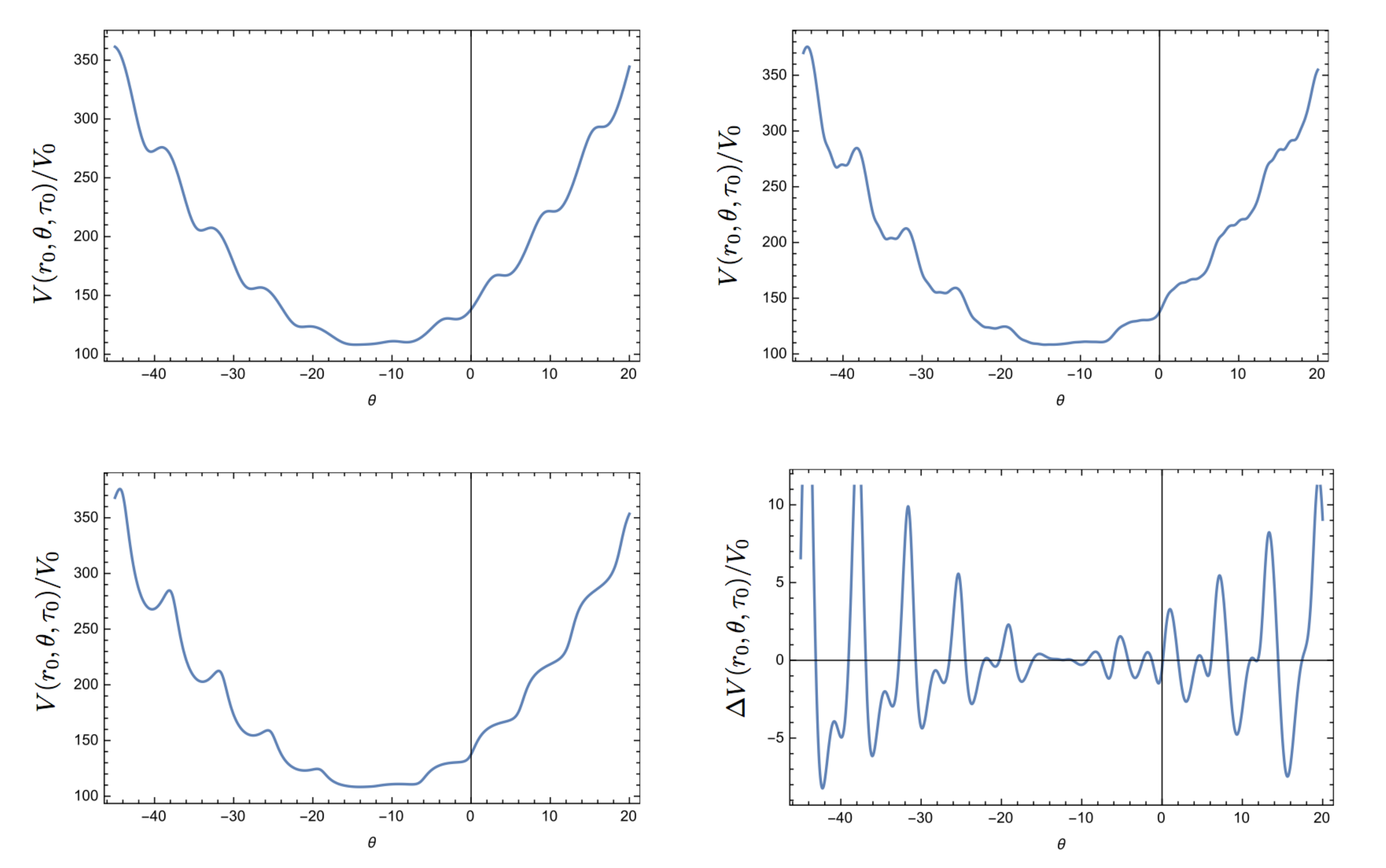}
\caption{
Scalar potential vs. $\theta$ and  the rest of the moduli fixed at their vev's for a configuration with non-zero fluxes $F_1=H_1=1$, $F_4=H_2=-10$, $F_2=F_3=0$, $H_3=H_4=1$ and a Minkowski vacuum. The values for the rest of the moduli are: $t_1=-6.28$,  $t_2=16$, $\theta_0=-12$, $r_0=0.4$.   The figures show the scalar potential  approximations to order 1, 4, 200 in $z$ from left to right and up to down. The last plot indicates the difference between order 200 and the order one calculation. The  $\theta$ asymmetry arises due to the odd powers of $\theta$ multiplying oscillatory functions appearing in the potential.  
Here again $V_0 = \alpha'^2/(2 \kappa_{10}^2 g_s)$  and the fluxes are given in string units. }
\label{mod1}
\end{center}
\end{figure}
\begin{figure}
\begin{center}
\includegraphics[width=.9\textwidth]{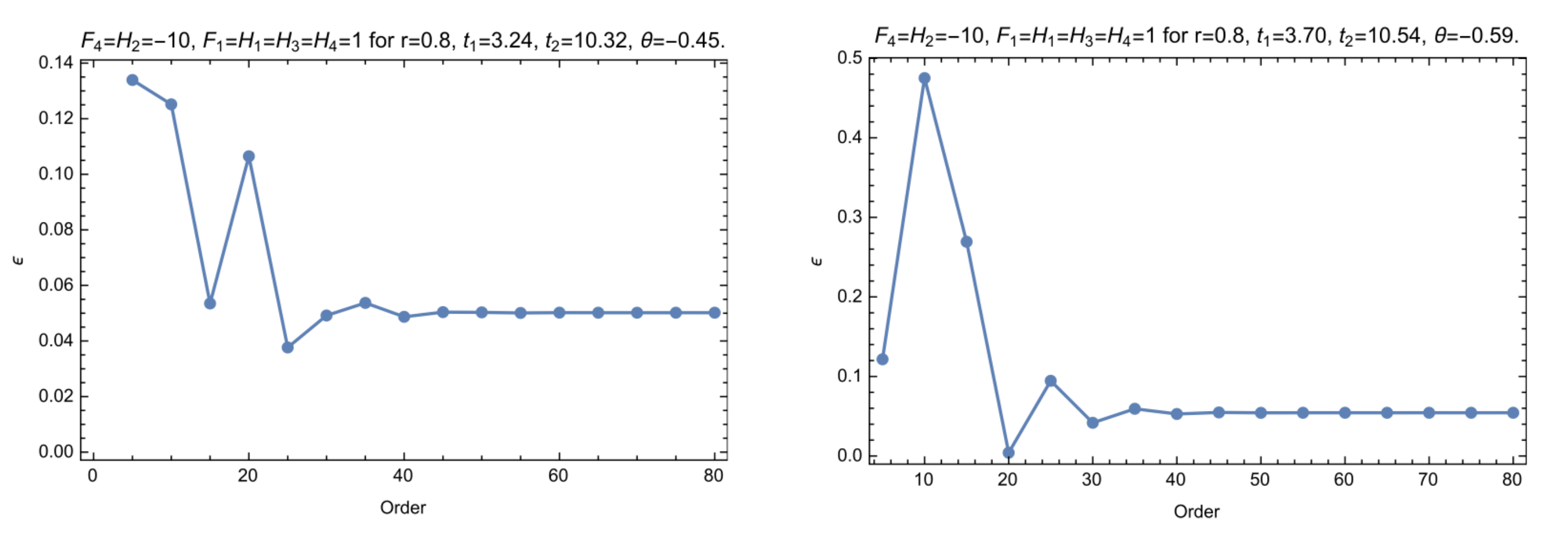}
\caption{The value of $\epsilon$  vs.~the order in the periods' series expansion.
The fluxes are the same that in Fig.\ref{mod-inf1}. The values of the moduli for each case are given above the figures. This shows that  $\epsilon$'s convergence is slow as we increase the order in the series expansion. The values of $\epsilon$ can differ in a $92 \%$ from the order 20 to the order 100. Convergence
is achieved: the value of $\epsilon$ at order 80 differs from the order 100 by $6\times 10^{-4}\%$. }
\label{mod-inf}
\end{center}
\end{figure}

\begin{figure}
\begin{center}
\includegraphics[width=.8\textwidth]{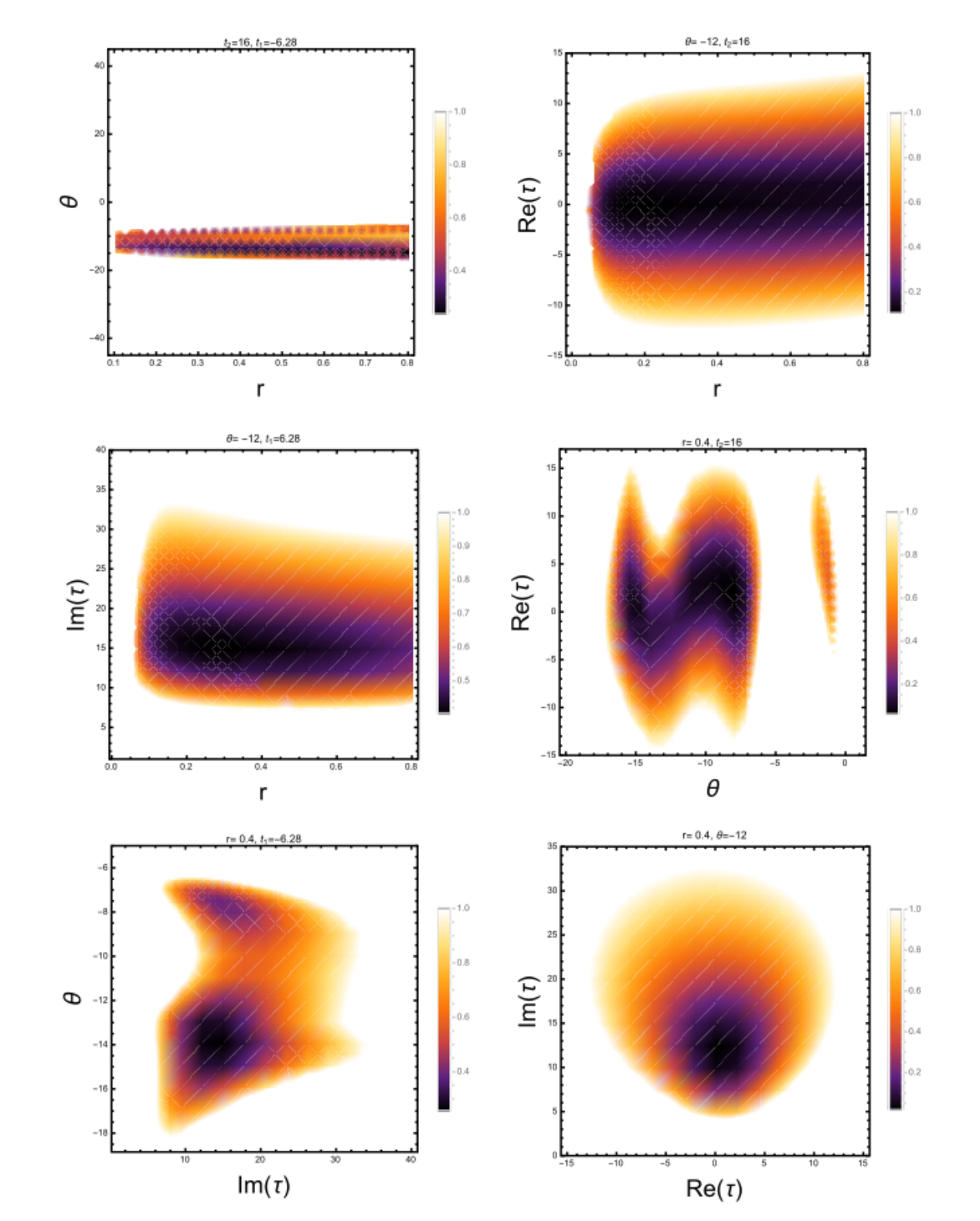}
\caption{  $\epsilon$ density plot for the flux configuration $F_1=H_1=H_3=H_4=1$, $F_4=H_2=-10$, $F_2=F_3=0$.
The CS modulus is in the conifold patch $z_C=r e^{i\theta}$ and evaluations are performed to order $200$ in the period series expansion $\Pi_C$.  The projections on the different planes  are made by fixing the   values of the moduli  at $t_1=-6.28,  t_2=16,  r=0.4,  \theta=-12$, respectively. 
The smaller values of $\epsilon$ in this region turned out be of order $\epsilon\sim0.05$. 
In this region we also find $\eta<1$. A sample $\eta$ eigenvalue is $\eta_{100}\sim -0.07$ for
the point $r =0.37 $, $t_1,\theta$ as before and  $t_2=9$, corresponding to the eigenvector $\sim (0.12, 0.004, 0.99, 0.09)$, which indicates that the $r$ direction is the dominant one along the inflationary direction. For the canonically normalized fields, the displacements in Planck units on the represented region are  of order 
$\Delta \phi_r \sim 0.1 M_{Pl}$, 
$\Delta \phi_{\theta}\sim 0.3 M_{Pl}$,
$\Delta \phi_{t_2}\sim 0.79 M_{Pl}$, $\Delta \phi_{t_1} \sim 0.78 M_{Pl}$. 
\label{mod-inf1}}
\end{center}
\end{figure}

\begin{figure}%
\begin{center}
\includegraphics[width=.99\textwidth]{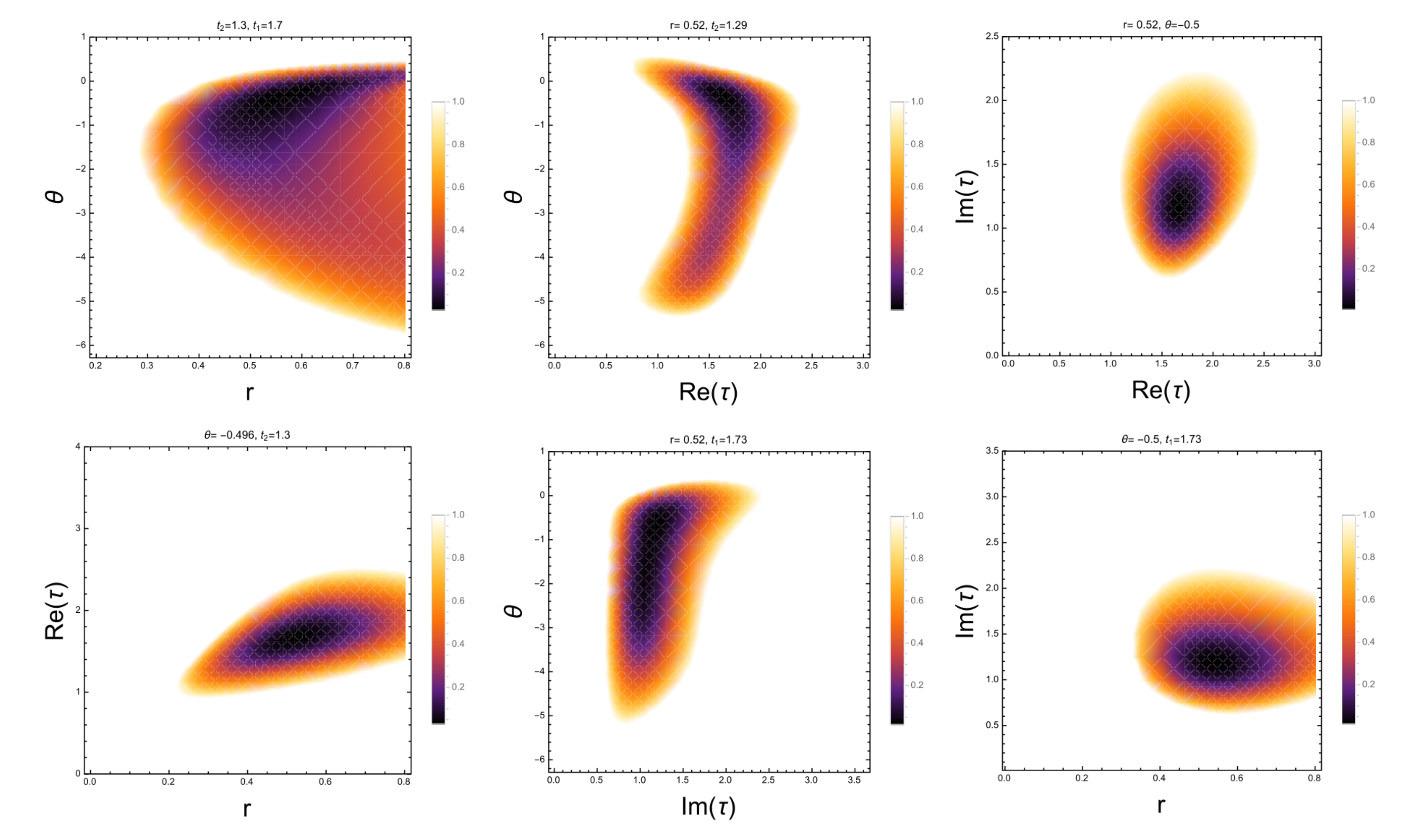}
\caption{
Density plot for $\epsilon$ for a configuration of non-zero fluxes $F_1=2$, $F_2=2$, $F_3=1$, $F_4=3$, $H_1=4$, $H_2=4$, $H_3=2$ and $H_4=1$. The CS modulus is in the conifold patch $z_C=r e^{i\theta}$ and evaluations are performed to order $200$ in the period series expansion $\Pi_C$. The projections on the different planes  are made by fixing the   values of the moduli  at 
$r=0.52$, $t_1=1.73$, $\theta= -0.496$, $t_2=1.295$ respectively. For this configuration, there is a $dS$ vacuum approximately at  $r=0.63, t_1=1.55, \theta=-0.03, t_2=1.25$. The smallest values of $\epsilon$ in these regions are $\epsilon\sim 0.03$. A sample $\eta$ eigenvalue is $\eta_{50}=-0.011$, the subindex denotes that  $\eta$ is computed at order $50$, corresponding to the eigenvector $\eta_{min}\sim (0.67, -0.71, -0.096, -0.22)$, giving as the preferential inflationary directions along $t_1$ and $t_2$.  For the canonically normalized fields the displacements in Planck units,
in the represented region, are of order
$\Delta \phi_{r}\sim 0.124 M_{Pl}$, 
$\Delta\phi_{\theta}\sim 0.58 M_{Pl}$,  
$\Delta \phi_{t_2}\sim 0.69 M_{Pl}$, 
$\Delta \phi_{t_1} \sim 0.65 M_{Pl}$.}
\label{eCon3}
\end{center}
\end{figure}

\begin{figure}%
\begin{center}
\includegraphics[width=.99\textwidth]{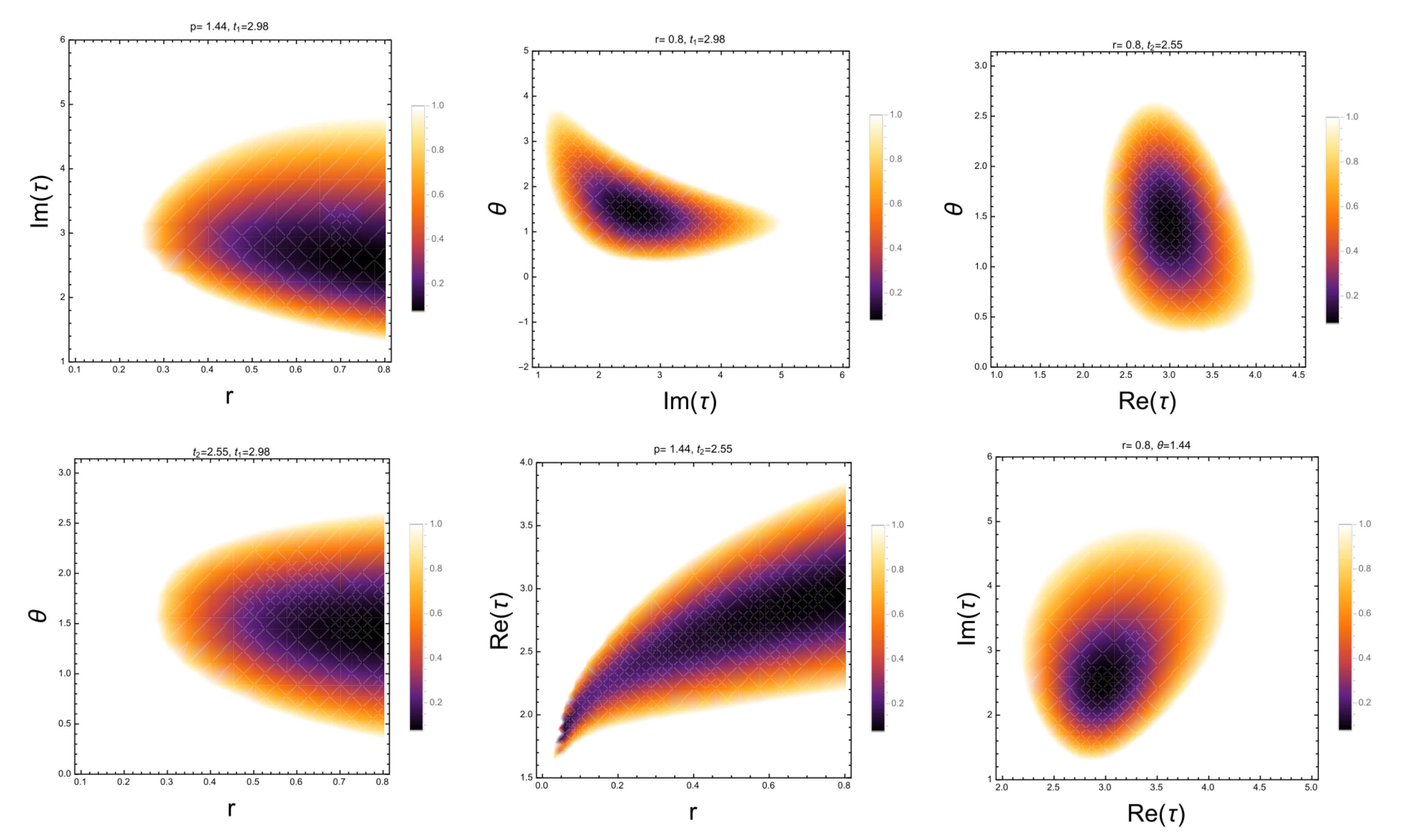}
\caption{ 
$\epsilon$ density plot on the six different planes for a configuration of non-zero fluxes $F_1=1$, $F_3=10$, $H_1=3$, $H_3=2$, $H_4=1$.  The CS modulus is in the conifold patch $z_C=r e^{i\theta}$ and evaluations are performed to order $200$ in the period series expansion $\Pi_C$.  The planes are defined by setting two
of the fields to $r_0=0.8, t_{1,0}=2.98, \theta_0=1.44, t_{2,0}=2.55$ respectively, at this
point  $\epsilon\sim 0.08$.  For the canonically normalized fields the displacements in Planck units 
of the represented region are of order  
$\Delta \phi_r \sim 0.045 M_{Pl}$,  
$\Delta \phi_{\theta}\sim 0.065 M_{Pl}$, 
$\Delta \phi_{t_2}\sim 0.8 M_{Pl}$, 
$\Delta \phi_{t_1} \sim 0.44M_{Pl}$.
There is a minimum $\eta$ eigenvalue $\eta_{100}=-0.025$ at the point $r=r_0, \theta=\theta_0, t_1=t_{1,0}, t_2=t_{2,0}$
with eigenvector $\sim (-0.98, 0.15, -0.15, 0.04)$ which signals $t_1$ as the dominant contribution along the inflationary direction on that point.}
\label{eCon1}
\end{center}
\end{figure}

\begin{figure}%
\begin{center}
\includegraphics[width=.99\textwidth]{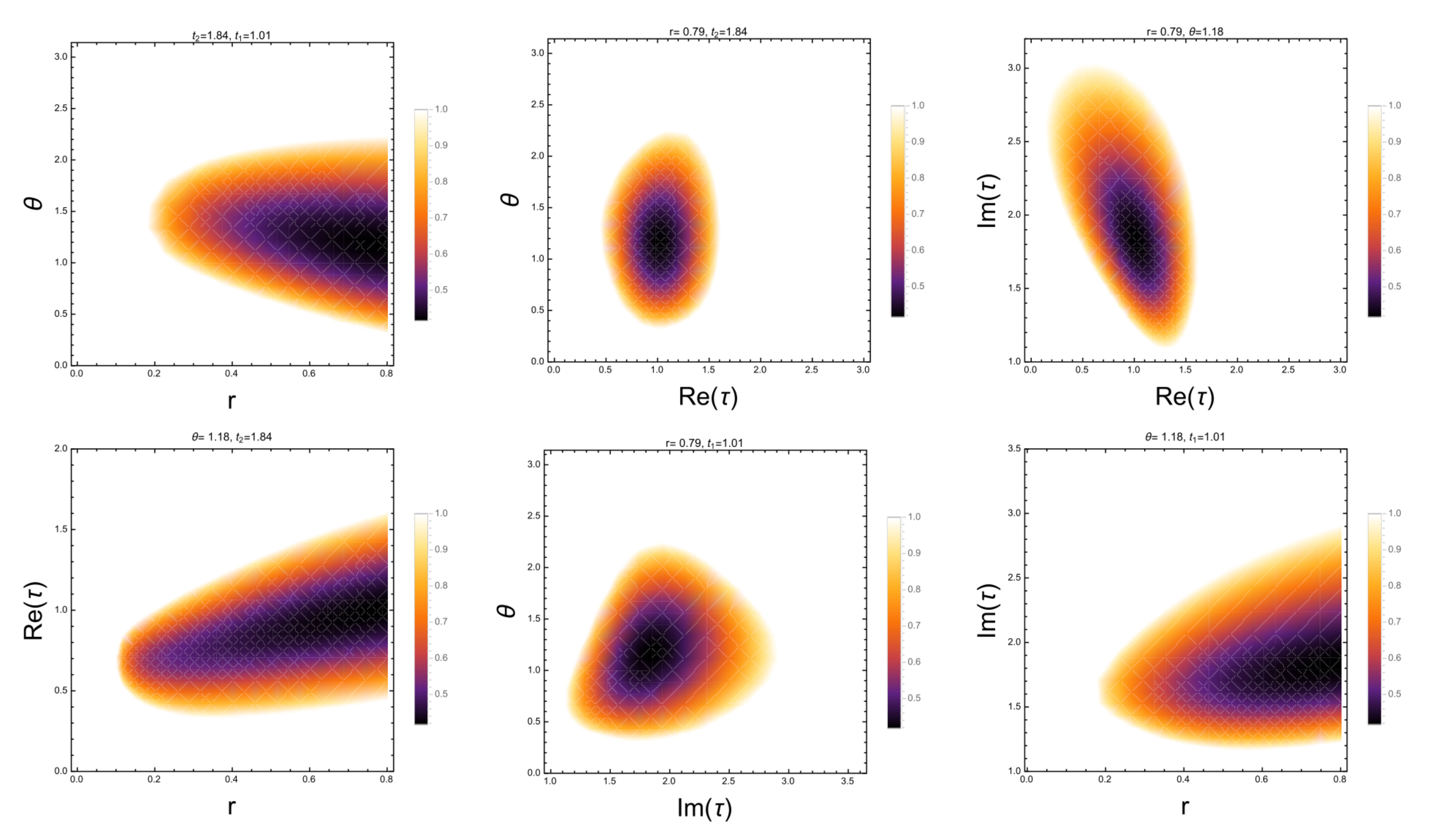}
\caption{ 
$\epsilon$ density plots  on the six planes of two variables for a configuration of non-zero fluxes $F_1=2$, $F_3=6$, $H_1=4$, $H_3=2$, $H_4=2$. The CS modulus is in the conifold patch $z_C=r e^{i\theta}$ and evaluations are performed to order $200$ in the period series expansion $\Pi_C$. 
At $r=0.79$, $t_{1}=1.01$, $\theta=1.18$, $t_{2}=1.84$ we have $\epsilon\sim 0.417$,
there are smaller values of $\epsilon$ in the orbifold convergence region.  
A sample $\eta$ eigenvalue inside the $\epsilon<1$ region is $\eta_{100}=-0.06$ at $r=0.18, t_1=0.75$ and $\theta=1.18, t_2=1.84$. The $\eta$ eigenvector $-(0.8,0.6,0.1,0.01)$ shows that at this point, the inflationary direction is mostly along  $t_1$ and $t_2$. For the canonically normalized fields the displacements in Planck units of the represented
region are of order 
$\Delta \phi_r \sim 0.055  M_{Pl}$,  
$\Delta \phi_{\theta}\sim 0.046  M_{Pl}$, 
$\Delta \phi_{t_2}\sim 0.49 M_{Pl}$, 
$\Delta \phi_{t_1} \sim 0.41 M_{Pl}$. }
\label{eCon2}
\end{center}
\end{figure}

\begin{figure}
\begin{center}
\includegraphics[width=.99\textwidth]{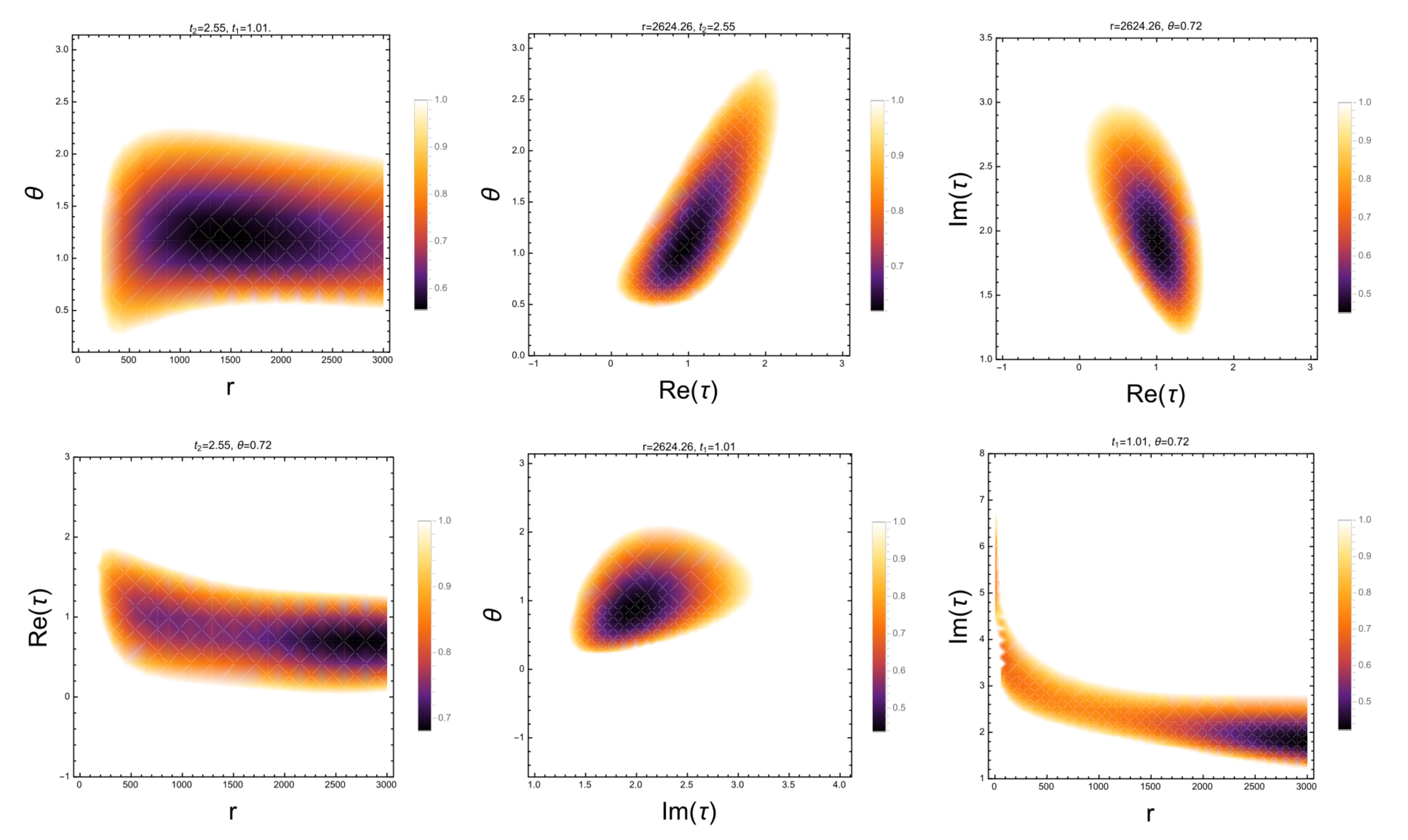}
\caption{ %
$\epsilon$ density plots  on the six planes for the same flux configuration as in Fig.~\ref{eCon2}  
 in the orbifold series convergence region. The CS modulus is in the orbifold patch $z_O=r e^{i\theta}$ and evaluations are performed to order $200$ in the period series expansion $\Pi_O$.  A sample minimum $\eta$ eigenvalue is -0.016,
for the point  $t_1^0 =1.010$, $\theta^0= 0.7178$, $t_2^0=3.68$ and $r^0=2000$ 
with eigenvector $\sim(0.8, 0.6, -0.00023, 0.3)$.
There is a saddle of the potential at $r =1135.59$, $p=-1.078$, $t_1=0.8071$, $t_2=1.625$, with eigenvector $\sim(0.036, 0.0355,0.002, -0.999)$ giving that the unstable direction is mostly $\theta$. 
For the canonically normalized fields the displacements in Planck units of the represented region are  
 $\Delta \phi_{r}\sim 0.037 M_{Pl}$,  
 $\Delta\phi_{\theta}\sim 0.19 M_{Pl}$,  
 $\Delta \phi_{t_2}\sim 0.49 M_{Pl}$,  
 $\Delta \phi_{t_1} \sim 0.29M_{Pl}$. \label{eOrb1}}
\end{center}
\end{figure}

\begin{figure}
\begin{center}
\includegraphics[width=.9\textwidth]{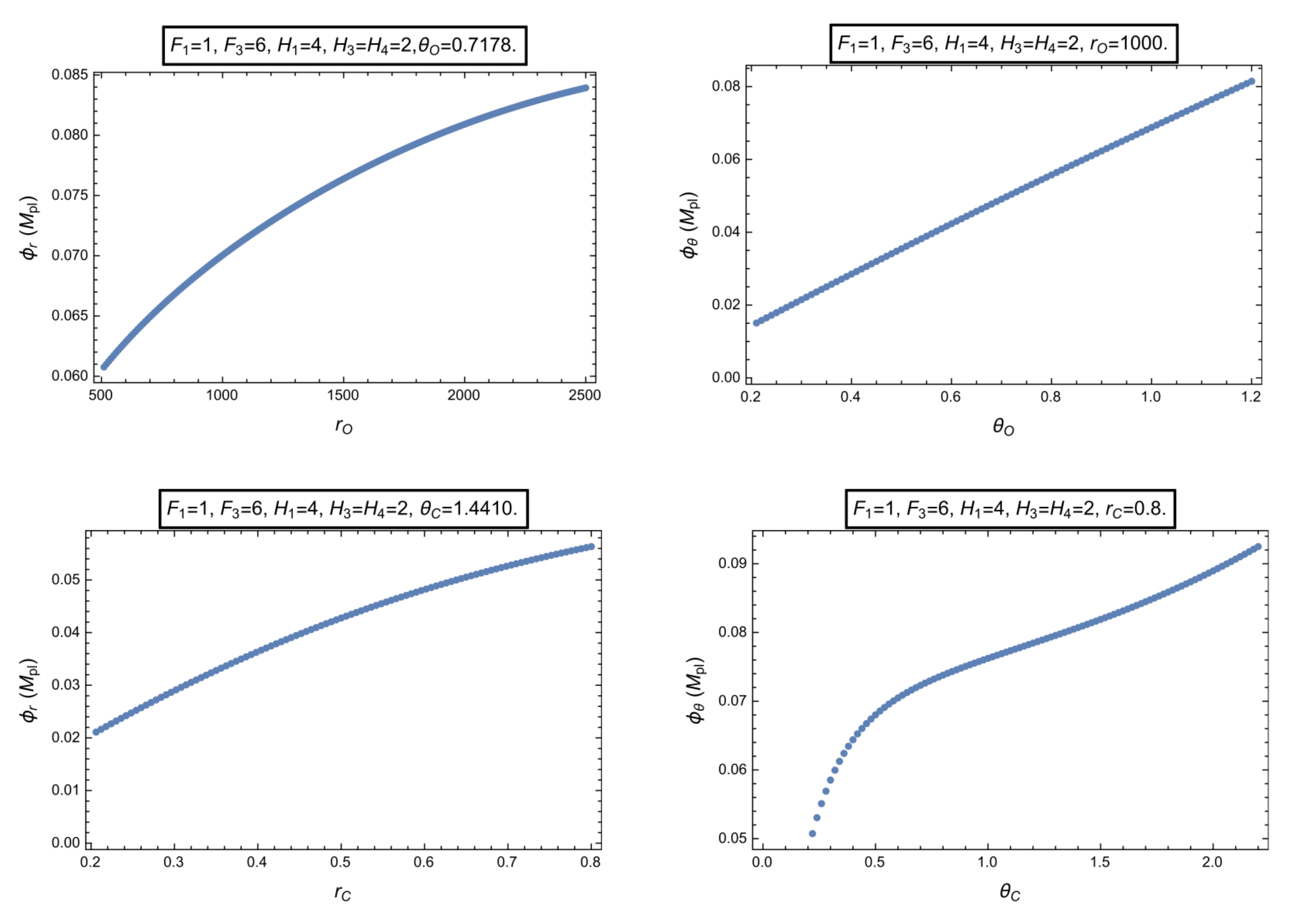}
\caption{ 
Trajectories of the locally  normalized fields for $r$ and $\theta$, $\phi_r$ and $\phi_{\theta}$ respectively, in the orbifold and conifold  convergence regions. 
The displacement of the canonically normalized field $\phi_r$ is always sub-Planckian,  whereas the displacement of the canonically normalized field $\phi_\theta$
can be arbitrarily large.
\label{Mpldisp}}
\end{center}
\end{figure}

\newpage

\section{Conclusions}
\label{conclusions}

We explored the moduli space of no-scale   type IIB orientifold flux compactification on the   mirror quintic Calabi-Yau 3-fold.  
For the complex structure modulus, we  solved the Picard-Fuchs equations in four convergence regions: the orbifold, the conifold, the large complex structure points patches and in a regular point patch.
This  allowed us  to have exact expressions for the periods in the  whole complex structure moduli space. 
The solutions to the PF equations have been previously studied in the literature \cite{Huang:2006hq, Candelas:1990qd}, 
and we have extended this study in the present  work by computing  them to all orders in  the series required  to achieve convergence.

Using these solutions  we explored the four dimensional  moduli space composed of the complex structure modulus $z$ and the axio-dilaton $\tau$. We  searched  for Minkowski vacua, vacua with hierarchies and regions with small multi-field slow-roll inflationary parameters $\epsilon, \eta$ (defined in eq.~\eqref{SRP}). 
We gave special attention to the periods in the conifold convergence region, where we  compared vacua  obtained using the series expansion to an arbitrary order approximation with those obtained using an approximation near the conifold point. 
We found that Minkowski vacua are in general absent  in the fundamental domain of
$\theta=\arg(z)$, while  vacua appear generically when monodromies around the conifold are taken.  We  pointed out  the importance of  considering higher order terms in the periods' series expansion in $z$ to ensure  the existence  of the vacua. Specifically, we found that some  Minkowski vacua appearing  at leading order near the conifold, disappear when
higher order terms in $z$ are considered. These fake  vacua turn out to be  an effect of the approximation,
and by careful analysis in a different patch (in the vicinity of the orbifold, LCS, or a regular point) they can be discarded.

We also found that for Minkowski vacua very close to the conifold point with the CS stabilised at $|z_0|\ll1$ and with $F_1,H_3,H_4$ being the only non-zero fluxes,    hierarchies in the physical scales exist  for large values of  $\frac{H_3}{H_4}$, in agreement with \cite{Giddings:2001yu}.  The vev of the axio-dilaton $\tau_0$ is fixed for given values of $F_1,H_4$  while $|z_0|$ decreases according to the value for  $H_3$, increasing the spacetime and compactification scales' hierarchy. 
For these flux configurations in Section \ref{hierarchies}  we find an extra term in the expression for the CS value at the minimum $z_0$, that changes
the hierarchy by an order of magnitude. The exact vacua differ from the approximated ones, but converge to them
when $\frac{H_3}{H_4}$ is increased.

In addition we studied general flux configurations, similar to those previously studied in  \cite{Ahlqvist:2010ki} with the near  conifold  approximation ($z_0 \ll1$).  We review their formula for $z_0$ and observe that $F_3$ can also be tuned to achieve a small value for $|z_0|$ while keeping $\tau_0$ constant. This provides 
a way of obtaining vacua with hierarchies between the 4D and 6D scales different from the ones in \cite{Giddings:2001yu}. Again it occurs that the actual vacua differ from the vacua obtained using the near-the-conifold approximation ($z_0 \ll1$),  but in the limit of highly negative $F_3$ they converge to the approximated  ones. In general it holds that for a generic flux configuration ensuring  that $|z_0|\ll1$ with $z_0$ in (\ref{aproxG}), the exact  and approximated vacua match.

For all flux configurations with a Minkowski vacuum, we did not find slow-roll inflationary regions of the scalar potential.
Numerically we  observed, through multiple searches, that having both, a vacuum solution near the conifold  and inflation regions, seems to be incompatible. 
This was done by   optimising a near  conifold vacuum $|z_0|<1$
and $\epsilon<1$ constraints varying the fluxes and the moduli. No 
physical solution with $g_s^{-1}= \Im(\tau_0)>1$ was found in the exploration.
 We further explored flux configurations with a Minkowski vacuum not necessarily closed to the conifold, but within the conifold convergence region. We looked for slow-roll regions along  $\theta = \arg(z)$ taking several monodromies
 w.r.t. the $\theta\in [0,2\pi)$ domain and keeping  the rest of moduli fixed, but this search turned out to be unsuccessful.  The modulations along $\theta$ of the scalar potential are highly affected by the power of $z$  considered in the periods' series expansions.  
We also found apparent flat regions of $\theta$, giving a small single field $\epsilon$, but
with an unstable radial direction.

We thus explored multi-field slow-roll regions, allowing all fields to evolve. The exploration was performed numerically,  considering the multi-field $\epsilon$ parameter and  varying all the moduli and the fluxes to optimise a minimum value of $\epsilon$. On the phase direction, monodromies affect modulations of the potential,  i.e. as $\theta$ grows the oscillations of $V$ also increase. This implies that  the only possible region with a flat potential  along the CS phase lies 
at the bottom of the scalar potential, within the basic region of $\theta$. 

We found flux configurations with inflationary regions going from the conifold to the orbifold patch.
In those regions the potential looks flat with small values of $\epsilon$ and small
minimum eigenvalues of the $\eta$-parameter.  In one of those cases a dS vacuum near the
inflationary region was found. This is interesting as it can serve as an explicit realisation of the uplift mechanism considered in \cite{SS} once we include stabilisation of the K\"ahler moduli, which could be addressed as in  \cite{DDFGK}.  We  computed the eigenvectors of $\eta$
in the regions with slow-roll and obtained that in general, the inflationary trajectory occurs  along  a combination of all the moduli directions. That is,  we did not find that axion monodromy inflation  mostly along the phase of $z$,  $\theta$ is realised.  
However, it is important to stress that we only looked for regions where the potential's slow-roll parameters \eqref{SRP} are small. It is possible that studying the more general slow-roll parameters for these type of bumpy potentials as recently considered in \cite{PTZ}, will give rise to successful inflation. 
Finally, we pointed out that the $\epsilon$ values in certain patches can depend strongly on the
approximation for the period series in $z$. Exact values of the periods
are required, since one might establish false conclusions from keeping only leading terms in $z$.

We showed that the total displacement for the canonically normalised fields in the $r$ ($r=r_O,r_C$ with $r_C=|z_C|$ or $r_O=|z_O|$) region (with the rest of moduli fixed) from the conifold to the orbifold is finite and smaller than a Planck unit.  
Thus an small excursion in Planck units of the $r$ canonical field may include any patch of the
CS moduli space. Therefore any study performed with $r$ bounded to a single convergence critical point region is 
inexact for models of inflation in the CS moduli space. On the other hand, the displacement along the canonically normalised $\theta$ direction can grow unconstrained (see Figure \ref{Mpldisp}).

Our  results  highlight  the importance  of considering
the exact solutions for the CY periods to explore vacua and cosmological applications of the
CS  and dilaton moduli potential. We have found that the vacua and  slow-roll conditions depend crucially on the approximation considered.
It is thus important in order to study phenomenological questions to keep all necessary terms in the period expansion 
until convergence is achieved. Our findings indicate that one needs to consider more general slow-roll inflationary solutions where the potential can have bumpy features and give distinctive inflationary predictions as discussed recently in \cite{PTZ}. This can be studied at a first stage in the no-scale approximation as considered here. However a more realistic scenario will have to take into account the stabilisation of the K\"ahler moduli, which is a step that  needs to be taken next. 
 In addition the study of F-theory models with general axio-dilaton profile could offer new interesting possibilities. 
 As an advantage the K\"ahler moduli \cite{Witten:1996bn}  together with the CS moduli \cite{Gukov:1999ya} can be incorporated in the effective action. 
  CY 4-fold with more than one modulus are required, and the analysis of periods in the whole CS moduli space \cite{Bizet:2014uua} could be useful to explore further the role of the monodromies 
in inflationary scenarios.

\subsection*{Acknowledgements}
We would like to thank Alejandro Cabo Bizet, Cesar Dami\'an, Shinji Hirano, Gabriel Lopez Cardoso, Gustavo Niz, Octavio Obreg\'on, Fernando Quevedo and Miguel Sabido for useful discussions. We would also like to thank Shusha Parameswaran and Gianmassimo Tasinato for useful discussions and comments on the manuscript. We would like to  thank  specially  Albrecht Klemm for many helpful discussions, suggestions and comments on the manuscript. NC-B thanks the support of the NRF of South Africa,  PROMEP and PNCB, CITMA. OL-B was partially supported by DAIP-UG project number 640/2015 and CONACYT project number  257919. The work of OL-B. and IZ was partly supported by a Royal Society Newton Mobility Grant under number NI150084. IZ thanks the Physics Department, Campus Le\'on, University of Guanajuato for hospitality during her visit.


\appendix

\section{Transition matrices and periods on the symplectic basis}
\label{transition1}

The transition matrices are computed by taking sample points lying at the intersections
of the orbifold-conifold and conifold-LCS regions.  We denote with $S$ the integral symplectic basis (\ref{Pi1lcs}).
$M_{X,S}$ denotes the transition matrix from the coordinates $X$ to the coordinates $S$.
$X=C, M, O$ denotes the conifold ($C$), LCS ($M$) and orbifold ($O$) bases for the 
solutions of the PF equations, $\pi_C$ (\ref{Pcon}), $\pi_M$ (\ref{Plcs}), and $\pi_O$ (\ref{Porb}).  
We denote the periods expansions around the conifold, the LCS and the orbifold points in the integral symplectic basis (\ref{Pi1lcs}) 
as $\Pi_C$, $\Pi_M$ and $\Pi_O$ respectively. The transition matrix from the solutions $\pi_M$ to the basis (\ref{Pi1lcs}) is given by
\begin{equation}
M_{M,S}=\left(
\begin{array}{cccc}
-\frac{i 200}{8\pi^3}\zeta(3)& \frac{50}{24}\frac{1}{2 \pi i}&0& \frac{1}{(2 \pi i)^3}\\
\frac{50}{24}& - \frac{11}{2}\frac{1}{2 \pi i}&-\frac{1}{(2 \pi i)^2}& 0\\
1&0&0&0\\
0&\frac{1}{2\pi i}&0&0
\end{array}
\right)
\end{equation}

The change of basis matrix from the conifold basis $\pi_C$ to the integral symplectic basis has six coefficients that can only be determined numerically, those are  $a,b,c,d,e$ and $g$ \cite{Huang:2006hq}. In our calculations we have determined the elements of the matrix with 40 digits of precision to be \footnote{The elements of $M_{C,S}$ given till order 20 are: $a=6.1950162771495748881$, $b=1.01660471670258207478$, $c=-0.14088997944883090936$, $d=1.0707258684301558006$, $e=-0.024707613804484718111$, 
$f=0.0057845115995744470969$, $g=1.2935739845041086377$, $h=0.15076669512354730097$, $r=-0.027792180016865244887$.}
\begin{equation}
M_{C,S}=\left(
\begin{array}{cccc}
0& - \frac{\sqrt{5}}{2\pi} i&0& 0\\
a - \frac{11}{2} i g&b - \frac{11}{2} i h&c - \frac{11}{2} i r&0\\
d&e& f &- \frac{\sqrt{5}}{(2\pi i)^2}\\
 i g& i h& i r&0
\end{array}
\right)
\end{equation}
The transition matrix $M_{O,S}$ between the orbifold basis and the integral symplectic basis $\Pi_O$ (\ref{Pi1lcs}) satisfying $\Pi_O=M_{O,S}\pi_O$ has an inverse
\begin{equation}
M_{O,S}^{-1}=\left(
\begin{array}{cccc}
-\frac{16 e^{\frac{2\pi i}{5}}\pi^4}{(  e^{\frac{2\pi i}{5}}-1) \Gamma[\frac{1}{5}]^5}&
-\frac{16 e^{\frac{2\pi i}{5}}\pi^4}{(  e^{\frac{2\pi i}{5}}-1)^2 \Gamma[\frac{1}{5}]^5}&
\frac{80 e^{\frac{2\pi i}{5}}(1-e^{\frac{2\pi i}{5}}+e^{\frac{4\pi i}{5}})\pi^4}{(  e^{\frac{2\pi i}{5}}-1)^4 \Gamma[\frac{1}{5}]^5}&
-\frac{16 e^{\frac{2\pi i}{5}}(-3+8 e^{\frac{2\pi i}{5}})\pi^4}{(  e^{\frac{2\pi i}{5}}-1)^3 \Gamma[\frac{1}{5}]^5}\\
-\frac{16 e^{\frac{4\pi i}{5}}\pi^4}{(  e^{\frac{4\pi i}{5}}-1) \Gamma[\frac{2}{5}]^5}&
-\frac{16 e^{\frac{4\pi i}{5}}\pi^4}{(  e^{\frac{4\pi i}{5}}-1)^2 \Gamma[\frac{2}{5}]^5}&
\frac{80 e^{\frac{4\pi i}{5}}(1+e^{-\frac{2\pi i}{5}}-e^{\frac{4\pi i}{5}})\pi^4}{(  e^{\frac{4\pi i}{5}}-1)^4 \Gamma[\frac{2}{5}]^5}&
-\frac{16 e^{\frac{4\pi i}{5}}(-3+8 e^{\frac{4\pi i}{5}})\pi^4}{(  e^{\frac{4\pi i}{5}}-1)^3 \Gamma[\frac{2}{5}]^5}\\
-\frac{32 e^{-\frac{4\pi i}{5}}\pi^4}{(  e^{\frac{-4\pi i}{5}}-1) \Gamma[\frac{3}{5}]^5}&
-\frac{32 e^{-\frac{4\pi i}{5}}\pi^4}{(  e^{\frac{-4\pi i}{5}}-1)^2 \Gamma[\frac{3}{5}]^5}&
\frac{160 e^{-\frac{4\pi i}{5}}(1+e^{\frac{2\pi i}{5}}-e^{-\frac{4\pi i}{5}})\pi^4}{(  e^{-\frac{4\pi i}{5}}-1)^4 \Gamma[\frac{3}{5}]^5}&
-\frac{32 e^{-\frac{4\pi i}{5}}(-3+8 e^{-\frac{4\pi i}{5}})\pi^4}{(  e^{-\frac{4\pi i}{5}}-1)^3 \Gamma[\frac{3}{5}]^5}\\
-\frac{96 e^{-\frac{2\pi i}{5}}\pi^4}{(  e^{-\frac{2\pi i}{5}}-1) \Gamma[\frac{4}{5}]^5}&
-\frac{96 e^{-\frac{2\pi i}{5}}\pi^4}{(  e^{-\frac{2\pi i}{5}}-1)^2 \Gamma[\frac{4}{5}]^5}&
\frac{480 e^{-\frac{2\pi i}{5}}(1-e^{-\frac{2\pi i}{5}}+e^{-\frac{4\pi i}{5}})\pi^4}{(  e^{-\frac{2\pi i}{5}}-1)^4 \Gamma[\frac{4}{5}]^5}&
-\frac{96 e^{-\frac{2\pi i}{5}}(-3+8 e^{-\frac{2\pi i}{5}})\pi^4}{(  e^{-\frac{2\pi i}{5}}-1)^3 \Gamma[\frac{4}{5}]^5}
\end{array}
\right)
\end{equation}
The transition matrix given numerically reads
\begin{equation}
M_{O,S}={\scriptsize \left(
\begin{array}{cccc}
0.587512i&- 0.171576 i&0.011701i&- 0.000103i\\
1.462844- 2.338201i&-0.038521+0.260823114 i&-0.002627 - 0.017787i&0.000256+ 0.000409i\\
0.404320- 0.293756i&-0.027874+ 0.085788i&-0.001901- 0.005850i&0.000071+ 0.000051i\\
0.425127i&- 0.047422i&0.003234i&- 0.000074i
\end{array}
\right)}
\end{equation}
We give the first three order of the periods vs. $z_C$ on the integral symplectic basis. This
expression is obtained by acting with $M_{C,S}$ on $\pi_C$ in (\ref{Pcon}).
The period vector $\Pi_C=(\Pi_{C,1},\Pi_{C,2},\Pi_{C,3},\Pi_{C,4})$ has components
\begin{eqnarray}\scriptsize
\Pi_{C,1}&=&- 0.355881 i z - 0.249117 i z^2 -0.194548 i z^3+O(z^4), \label{PiC}\\
\Pi_{C,2}&=&6.19502 -7.11466 i + (1.0166 - 0.829217 i) z + (0.570733 - 0.427595 i) z^2 \nn \\
&+& (0.401804 - 0.287548 i) z^3+O(z^4),\nn\\
\Pi_{C,3}&=&1.07073 +\alpha z - 0.011511 z^2 - 0.006565z^3\nn \\
&-& \frac{\ln z}{2\pi i}(-2 \pi i\beta z - 0.249117 i z^2 -0.194548 i z^3)+O(z^4),\nn \\
\Pi_{C,4}&=&1.29357i+ 0.150767i z +0.0777445 i z^2 + 0.0522815 i z^3+O(z^4). \nn
\end{eqnarray}
Let us define here the coefficients as $\alpha=-0.024708$ and $\beta=0.056640$,
these are employed in Section \ref{hierarchies}. 
Observe that the monodromy is explicit because $\Pi_C^3=-\Pi^1_C \ln z/(2\pi i)+Q(z)$. 
The periods vs. $z_O$ on the integral symplectic basis are 
obtained by  $\Pi_O=M_{O,S} \pi_O$ in (\ref{Porb}).
The period vector $\Pi_O=(\Pi_{O,1},\Pi_{O,2},\Pi_{O,3},\Pi_{O,4})$ has components

\begin{eqnarray}
\Pi_{O,1}&=&0.587512 i z^{1/5} -0.171576 i z^{2/5} + 0.0117008 i z^{3/5}+O(z^{4/5}),\label{PiO}\\
\Pi_{O,2}&=&(1.46284 - 2.3382 i) z^{1/5} - (0.0385212 - 0.260823 i) z^{2/5}\nn \\
&-& (0.002627 + 0.0177871 i) z^{3/5}+O(z^{4/5}),\nn \\
\Pi_{O,3}&=&(0.40432 - 0.293756 i) z^{1/5} - (0.0278742 - 0.0857879 i) z^{2/5}\nn \\
&-& (0.00190091 + 0.00585041 i) z^{3/5}+O(z^{4/5}),\nn\\
\Pi_{O,4}&=&0.425127 i z^{1/5} -0.0474224 i z^{2/5} +  0.00323402 i z^{3/5}+O(z^{4/5}).\nn
\end{eqnarray}
The periods vs. $z_M$ on the integral symplectic basis are obtained by acting with $M_{M,S}$ on $\pi_M$ in (\ref{Plcs}). To obtain $\Pi_M=(\Pi_{M,1},\Pi_{M,2},\Pi_{M,3},\Pi_{M,4})$ with
\begin{eqnarray}
\Pi_{M,1}&=&-\frac{25 i \zeta(3)}{\pi^3}+\left(-\frac{2875 i}{4 \pi^3} - \frac{9625 i}{2 \pi} - \frac{3000 i \zeta(3)}{\pi^3}\right)z+
\left(-\frac{16491875 i}{32 \pi^3} - \frac{6751875 i}{8 \pi} -
\frac{2835000 i \zeta(3)}{\pi^3}\right) z^2 \nn\\
&+& \ln z\left(- \frac{25}{24 \pi}i +\left(\frac{2875}{8 \pi^3}- \frac{125}{\pi}\right) i z +\left(\frac{21040875}{32\pi^3} - \frac{118125}{\pi}\right)i z^2 \right)\nn \\
&+&\frac{5 i}{16 \pi^3}\ln z^2 (770 z+810225 z^2)+\frac{i}{48 \pi^3} \ln z^3(1+5! z+\frac{10!}{2^5} z^2 )+O(z^3),\nn\\
\Pi_{M,2}&=&\frac{25}{12}+\left(250 + \frac{2875}{4 \pi^2} + \frac{4235 i}{2 \pi}\right)z+\left(236250 + \frac{21040875}{16 \pi^2} + \frac{8912475 i}{4 \pi}\right)z^2,\label{PiM} \\
&+& \ln z \left(\frac{11 i}{4 \pi} +z\left( \frac{1925}{2 \pi^2} + \frac{330 i}{\pi} \right)+ z^2\left(\frac{4051125}{4 \pi^2} + \frac{311850 i}{\pi}\right)\right)+\nn \\
&+&  \frac{5}{8\pi^2}\ln z^2 \left(1+5! z+\frac{10!}{2^5}z^2\right)+O(z^3),\nn\\
\Pi_{M,3}&=&1+5! z+\frac{10!}{2^5} z^2+O(z^3),\nn \\
\Pi_{M,4}&=&\frac{1}{2\pi i}(770 z+810225 z^2)z+\frac{\ln z}{2\pi i}(1+5! z+\frac{10!}{2^5}) z^2+O(z^3).\nn
\end{eqnarray}

\section{Hierarchies}
\label{Ahierar}

Let us 
summarise the correction to the hierarchy formula of \cite{Giddings:2001yu} using the  notation of that paper.  The non-zero fluxes in their notation are
$M,K,K'$ which we denote as $F_1,H_3,H_4$. The four components of the periods are described on those coordinates as
\begin{equation}
\Pi(z)= (z,z'(z),\mathcal{G}(z),\mathcal{G}'(z)),
\end{equation}
while we note them as $\Pi_i$. The third component of the period vector has in general the properties \cite{Giddings:2001yu}
\begin{equation}
\mathcal{G}(z)=\frac{z\ln z}{2\pi i}+\text{hol.},\ \mathcal{G}(0)\neq 0,\ \ \ \partial_z{\mathcal{G}}(z)=\frac{\ln z}{2\pi i}+\delta_1(z).
\end{equation}
Closed to the conifold when $z\rightarrow 0$ we evaluate the following quantities
\begin{eqnarray}
\bar\Pi_0 \Sigma \Pi_0&=&\bar z'(0)\mathcal{G}'(0)-\bar{\mathcal{G}}'(0) z'(0),\nn \\ \bar\Pi_0 \Sigma \partial_z\Pi_0&=&\bar{\mathcal{G}}(0)-\bar z'(0)\partial_z{\mathcal{G}}'(0)-\bar{\mathcal{G}}'(0) \partial_z{z}'(0),\nn\\
\tau_0&=&\frac{M\bar{\mathcal{G}}(0)}{K' \bar z'(0)},\nn \\
\partial_z K_0&=&\frac{\bar\Pi_0 \Sigma \partial_z\Pi_0}{ \bar\Pi_0 \Sigma \Pi_0},\nn\\
W_0&=&M\mathcal{G}(0)-\tau_0 K' z'(0).\nn
\end{eqnarray}
This will give a covariant derivate of $W$
\begin{eqnarray}
D_z W&=&M\left(\frac{\ln z}{2\pi i}+\delta_1(0)\right)-\tau_0(K+K'\dot{z}'(0))+\partial_z K_0 W_0+O(z),\\
&=&M\frac{\ln z}{2\pi i}-\tau_0 K+b_0+O(z),\nn \\
b_0&=&M\delta_1(0)-\tau_0 K'\dot{z}'(0)+\partial_z K_0 W_0,\nn\\
\Delta_0&=&b_0/M=\delta_1(0)-\frac{\bar{\mathcal{G}}(0)}{\bar z'(0)}\partial_z {z}'(0)+\partial_z K_0\left(\mathcal{G}(0)-\frac{\bar{\mathcal{G}}(0)}{\bar z'(0)}z'(0)\right).\nn 
\end{eqnarray}
This translates in that there is a hierarchy given by
\begin{eqnarray}
z_0&\sim& \exp2\pi i \left( \frac{\tau_0 K}{M}-\Delta_0\right),\\
&\sim& \exp2\pi i \left( \frac{K \bar{\mathcal{G}}(0)}{K' \bar z'(0)}-\Delta_0\right).\nn
\end{eqnarray}
The previous formula adds a factor  $\exp \left(-2\pi i \Delta_0\right)$ w.r.t. to
(3.18) in \cite{Giddings:2001yu} this constitutes a correction to the hierarchy between the 4D and 6D scales, 
which is independent of the fluxes.

\section{Effect of monodromies on the scalar potential}
\label{appC}

The existence of inflationary regions due to monodromies has been widely explored in the last years. A logical strategy is to look for Minkowski vacua (especially in no-scale models as in the present case) and move around the moduli vevs in order to find flat regions in the potential.  In Section \ref{inflation} we looked  at the profile of the scalar potential in the complex structure phase direction $\theta$, while keeping all other moduli  frozen at their vevs. We found that  the amplitude of oscillations in the scalar potential along $\theta$ increased as we encircled the conifold.
Here we present an alternative way of moving  away from the Minkowski minimum in $z$ along the direction determined by $D_\tau W=0$, implying a transformation of the axio-dilaton in the trajectory. 

In the examples studied and presented in Section~\ref{inflation}, we saw that moving away from the vacuum in the $\theta$-direction, the scalar potential starts oscillating and  the amplitude of the oscillations are not constant as would be expected from an effective potential of the form
\begin{equation}
V \sim \Lambda \sin\left(\frac{\theta}{f}\right).
\end{equation}
Here we present an analytical description of this feature. 
For this we  study  a displacement in the direction defined by   $D_\tau W=0$ while performing monodromies $z\rightarrow e^{2\pi i n}z, n\in \mathbb{Z}$ around the conifold.   This is a good approximation for small values of $g_s$ and $\tau$ fixed at its value at the Minkowski minimum. However we'll see that far from the vacuum, $g_s$ starts growing leading the potential to an unphysical region. Under n-monodromies the superpotential $W$ transforms as
\begin{equation}
W \rightarrow W -nG_1\Pi_1\equiv  W -n g,
\end{equation}
and by keeping $D_\tau W=0$ after monodromies, the dilaton transforms  as
\begin{equation}
\tau_0\rightarrow \tau '=\frac{\tau_0 \bar\H-n\bar{f}}{\bar{\H}-n\bar{h}},
\label{tau}
\end{equation}
where $\H= H^\dagger\Sigma\bar{\Pi}$, $f=F_1\Pi_1$ and $h=H_1\Pi_1$. Observe  that it is not possible to perform a $SL(2,\mathbb{Z})$ transformation of $\tau$ while keeping $D_\tau  W=0$. i.e. the $\tau$-transformation given by (\ref{tau}) is not an $SL(2,\mathbb{Z})$-transformation.

The transformed superpotential, over a region on which $D_\tau W=0$, is given by
\begin{eqnarray}
W \rightarrow &&W(\tau')-ng(\tau ')\nonumber\\
&=&\omega(n, \bar{Z_0})\left[W(\tau_0,z_0)-\frac{n}{\bar{\H}}\left(\bar{h}\F-\bar{f}\H+\bar{\H}g(\tau_0)\right)\right]\nonumber\\
&\equiv&\omega (W_0+\delta W)
\end{eqnarray}
where 
\begin{equation}
\omega(n,\bar{z_0})=\frac{\bar{\H}}{\bar{\H}-n\bar{h}},
\end{equation}
and $\F=F^\dagger\Sigma\Pi$.
For each  value of  $n$, the values of $W$ and the rest of terms ($\H, \F, h, f$) are fixed at the minimum point $z_0$ and $\tau_0$. Now, notice that since $D_zW_0=0$ and $D_\tau W=0$ in $\tau=\tau'$ we have that
\begin{equation}
V=e^{\K(z_0, \tau'(n))}\left|D_z(\omega \delta W)\right|^2_{z_0}K^{z\bar{z}}(z_0, \bar{z_0}),
\end{equation}
from which we can read that the only  contribution to the scalar potential comes from the complex structure's K\"ahler derivative. The final expression for $V$ is
\begin{equation}
V=\frac{ig_s(n)}{2}\frac{1}{\Pi^\dag\Sigma\bar{\Pi}(z_0)}\left|D_z\xi\right|^2_{z_0}K^{z\bar{z}}_0\left|\frac{n}{\bar{\H}_0-n\bar{h}_0}\right|^2,
\end{equation}
where 
\begin{equation}
\xi=2i\Im(f\bar{\H}-h\bar{\F}),
\end{equation}
and
\begin{equation}
\frac{1}{g_s(n)}=\Im~\frac{|\H|^2\bar{\F}\H+n\left(\bar{\H}^2\F h-\bar{f}\H|\H|^2\right)}{|\H|^2|\bar{\H}-n\bar{h}|^2}.
\end{equation}
Notice the following:
\begin{enumerate}
\item
For large $n$, $g_s$ tends to infinite (see also (\ref{tau})), implying that our perturbative analysis is only valid from $n=0$ to some finite $n$. Actually for $n$ satisfying
\begin{equation}
 |\H|^4n^2+\left[-2|\H|^2~\Re(\H\bar{h})+\Im(\bar{\H}^2\F h-\H^2\bar{\H}f)\right]n+|\H|^2\left[|h|^2-\Im(\bar{\F}\H)\right]<0,
\end{equation}
at $z=z_0$, for which $g_s<1$.
\item
From relation (\ref{tau}), it is straightforward to see that positive monodromies will lead to a string coupling with the opposite sign with respect to that at the minimum. Hence, if we start in a perturbative regime, for some positive $n$, we shall leave the physical region. This however does not happen for negative monodromies.
\end{enumerate}
By substituting the value of $g_s(n)$ in $V$, we get
\begin{equation}
V= \frac{n^2C}{A+Bn}(z_0)
\end{equation}
with $n$ negative and $A$, $B$ and $C$ functions valued at $z_0$ given by
\begin{eqnarray}
C(z_0)&=& \frac{i|\H|^2}{\Pi^\dagger\Sigma\bar{\Pi}}\left|D_z\xi\right|^2_{z_0}K^{z\bar{z}}_0,\nonumber\\
B(z_0)&=&\bar{\H}^2\F h-\bar{f}\H|\H|^2,\nonumber\\
A(z_0)&=&2~\Im(|\H|^2\bar{\F}\H).
\end{eqnarray}
These values of the scalar potential at entire displacements of $\theta$, show that for each shift on the phase $\theta\rightarrow \theta+2\pi$, the scalar potential increases its value. The analysis breaks down as $g_s$ becomes larger than unity or the scalar potential becomes negative, showing that our assumption on $D_\tau W=0$ cannot be kept on the whole physical moduli space. Between each consecutive value of $n$ the scalar potential oscillates with a minimum for some value of $\theta$  in the interval $\big[\theta+ 2\pi n, \theta + 2\pi(n+1)\big)$. This shows that for the given trajectory in the moduli space, one  expects an increase in the oscillations' amplitude of the scalar potential.

\bibliographystyle{utphys}
\bibliography{biblioV}

\end{document}